\documentclass[preprint,prd,aps,showpacs,showkeys,nofootinbib]{revtex4}
\usepackage{graphicx}
\begin{document}


\title{Two loop electroweak corrections to $\bar B\rightarrow X_s\gamma$ and $B_s^0\rightarrow \mu^+\mu^-$ in the B-LSSM}

\author{Jin-Lei Yang$^{a}$\footnote{JLYangJL@163.com},
Tai-Fu Feng$^{a}$\footnote{fengtf@hbu.edu.cn},
Hai-Bin Zhang$^{a}$\footnote{hbzhang@hbu.edu.cn}, Rong-Fei Zhu$^{a}$, Shu-Min Zhao$^{a}$, Xiu-Yi Yang$^{b}$}

\affiliation{$^a$Department of Physics, Hebei University, Baoding, 071002, China\\
$^b$Department of Science, University of Science and Technology Liaoning, Anshan, 114051, China}

\begin{abstract}
The rare decays $\bar B\rightarrow X_s\gamma$ and $B_s^0\rightarrow \mu^+\mu^-$ are important to research new physics beyond standard model. In this work, we investigate two loop electroweak corrections to $\bar B\rightarrow X_s\gamma$ and $B_s^0\rightarrow \mu^+\mu^-$ in the minimal supersymmetric extension of the SM with local $B-L$ gauge symmetry (B-LSSM), under a minimal flavor violating assumption for the soft breaking terms. In this framework, new particles and new definition of squarks can affect the theoretical predictions of these two processes, with respect to the MSSM. Considering the constraints from updated experimental data, the numerical results show that the B-LSSM can fit the experimental data for the branching ratios of $\bar B\rightarrow X_s\gamma$ and $B_s^0\rightarrow \mu^+\mu^-$. The results of the rare decays also further constrain the parameter space of the B-LSSM.

\end{abstract}


\maketitle

\section{Introduction\label{sec1}}
\indent\indent
The study on B physics is one of the most promising windows to detect the new physics beyond the standard model (SM), since the theoretical evaluations on the relevant physical quantities are not seriously affected by the uncertainties due to the QCD effects. Recently, the average experimental data on the branching ratios of $\bar B\rightarrow X_s\gamma$ and $B_s^0\rightarrow \mu^+\mu^-$ are shown as~\cite{17,BB1,BB2,BB3}
\begin{eqnarray}
&&Br(\bar B\rightarrow X_s \gamma)=(3.49\pm0.19)\times 10^{-4},\nonumber\\
&&Br(B_s^0\rightarrow \mu^+\mu^-)=(2.9_{-0.6}^{+0.7})\times10^{-9}.
\label{experimental data}
\end{eqnarray}
The SM predicts the $\bar B\rightarrow X_s\gamma$ and $B_s^0\rightarrow \mu^+\mu^-$ branching ratios to be~\cite{SMP1,SMP2,SMP3,SMP4,SMP5,SMP6,SMP7,SMP8,SMP9}
\begin{eqnarray}
&&Br(\bar B\rightarrow X_s \gamma)=(3.36\pm0.23)\times 10^{-4},\nonumber\\
&&Br(B_s^0\rightarrow \mu^+\mu^-)=(3.23\pm0.27)\times10^{-9},
\end{eqnarray}
which are in agreement with the experimental results very well. So, the precise measurements on the rare B-decay processes constrain the new physics beyond SM strictly.

In extensions of the SM, the supersymmetry is considered as one of the most plausible candidates. Actually, the analyses of constraints on parameters in the minimal supersymmetric extension of the SM (MSSM) are discussed in detail~\cite{B4,B5,B6,B7,B8,B9,B10,B11}. The authors of Refs.~\cite{NPB1,NPB2,NPB3} present the calculation of the rate inclusive decay $B\rightarrow X_s\gamma$ in the two-Higgs doublet model (THDM). The supersymmetric effect on $B\rightarrow X_s\gamma$ is discussed in Refs.~\cite{NPB4,NPB5,NPB6,NPB8,NPB7,Zhang1,Feng1} and the next-to-leading order (NLO) QCD corrections are given in Ref.~\cite{NPB9}. The branching ratio for $B_s\rightarrow l^+l^-$ in THDM and supersymmetric extensions of the SM has been calculated in Refs.~\cite{He:1988tf,Skiba:1992mg,Choudhury:1998ze,Huang:2000sm,Feng2,Feng3}. The hadronic B decays~\cite{NPB11} and CP-violation in these processes~\cite{NPB12} have also been discussed. The authors of Ref.~\cite{NPB13} have discussed possibility of observing supersymmetric effects in rare decays $B\rightarrow X_s\gamma$ and $B\rightarrow X_s e^+ e^-$ at the B-factory. The supersymmetric effects on these processes are very interesting and studies on them may shed some light on the general characteristics of the supersymmetric model. A relevant review can be found in Refs.~\cite{NPB16,NPB17}.

The minimal supersymmetric extension of the SM with local $B-L$ gauge symmetry (B-LSSM)~\cite{5,6} is based on the gauge symmetry group $SU(3)_C\otimes SU(2)_L\otimes U(1)_Y\otimes U(1)_{B-L}$, where $B$ stands for the baryon number and $L$ stands for the lepton number respectively. Besides accounting elegantly for the existence and smallness of the left-handed neutrino masses, the B-LSSM also alleviates the aforementioned little hierarchy problem of the MSSM~\cite{search}, because the exotic singlet Higgs and right-handed (s)neutrinos~\cite{77,88,9,99,10,11,last1,last2,last3} release additional parameter space from the LEP, Tevatron and LHC constraints. The invariance under $U(1)_{B-L}$ gauge group imposes the the R-parity conservation which is assumed in the MSSM to avoid proton decay. And R-parity conservation can be maintained if $U(1)_{B-L}$ symmetry is broken spontaneously~\cite{C.S.A}. Furthermore, it could help to understand the origin of R-parity and its possible spontaneous violation in the supersymmetric models~\cite{S.K,P.F,V.B} as well as the mechanism of leptogenesis~\cite{J.P,K.S.B}. Moreover, the model can provide much more candidates for the Dark Matter comparing that with the MSSM~\cite{16,1616,DelleRose:2017ukx,DelleRose:2017uas}. In this work, we analyze two loop electroweak corrections to $\bar B\rightarrow X_s\gamma$ and $B_s^0\rightarrow \mu^+\mu^-$ in the B-LSSM. In this framework, new couplings and particles make new contributions to both of these processes with respect to the MSSM. The numerical results of the rare decays also further constrain the parameter space of the model.

Our presentation is organized as follows. In Sec. II, the main ingredients of B-LSSM are summarized briefly, including the superpotential, the general soft breaking terms, the Higgs sector and so on. Sec. III contains the effective Hamilton for $\bar B\rightarrow X_s\gamma$ and $B_s^0\rightarrow \mu^+\mu^-$. The numerical analyses are given in Sec. IV, and Sec. V gives a summary. The tedious formulae are collected in Appendices.

\section{The B-LSSM\label{sec2}}
\indent\indent
In the B-LSSM, one enlarges the local gauge group of the SM to $SU(3)_C\otimes SU(2)_L\otimes U(1)_Y\otimes U(1)_{B-L}$, where the $U(1)_{B-L}$ can be spontaneously broken by the chiral singlet superfields $\hat{\eta}_1$ and $\hat{\eta}_2$. In literatures there are several popular versions of B-LSSM. Here we adopt the version described in Refs.~\cite{44,8,Khalil:2015wua,Hammad:2016trm} to proceed our analysis, because this version of B-LSSM is encoded in SARAH~\cite{164,165,166,167,168} which is used to create the mass matrices and interaction vertexes of the model. Besides the superfields of the MSSM, the exotic superfields of the B-LSSM are three generations right-handed neutrinos $\hat{\nu}_i^c\sim$(1, 1, 0, 1) and two chiral singlet superfields $\hat{\eta}_{1}\sim(1,1,0,-1)$, $\hat{\eta}_{2}\sim(1,1,0,1)$. Meanwhile, quantum numbers of the matter chiral superfields for quarks and leptons are given by
\begin{eqnarray}
&&\hat{Q}_i=\left(\begin{array}{c}\hat U_i\\ \hat D_i\end{array}\right)\sim(3, 2, 1/6, 1/6), \quad\;\hat{L}_i=\left(\begin{array}{c}\hat \nu_i\\ \hat E_i\end{array}\right)\sim(1, 2, -1/2, -1/2),\nonumber\\
&&\hat{U}^c_i\sim(3, 1, -2/3, -1/6),\quad\; \hat{D}^c_i\sim(3, 1, 1/3, -1/6), \quad\;\hat{E}^c_i\sim(1, 1, 1, 1/2),
\end{eqnarray}
with $i=1,2,3$ denoting the index of generation. In addition, the quantum numbers of two Higgs doublets is assigned as
\begin{eqnarray}
&&\hat{H_1}=\left(\begin{array}{c}H_1^1\\ H_1^2\end{array}\right)\sim (1, 2, -1/2, 0),\quad\;\hat{H_2}=\left(\begin{array}{c}H_2^1\\ H_2^2\end{array}\right)\sim(1, 2, 1/2, 0).
\end{eqnarray}

The corresponding superpotential of the B-LSSM is written as
\begin{eqnarray}
&&W=W_{MSSM}+W_{(B-L)}.
\end{eqnarray}
Here, $W_{MSSM}$ is the superpotential of the MSSM, and $W_{(B-L)}$ is the sector involving exotic superfields,
\begin{eqnarray}
&&W_{(B-L)}=Y_{\nu, ij}\hat{L_i}\hat{H_2}\hat{\nu}^c_j-\mu' \hat{\eta}_1 \hat{\eta}_2
+Y_{x, ij} \hat{\nu}_i^c \hat{\eta}_1 \hat{\nu}_j^c,
\end{eqnarray}
where $i, j$ are generation indices. Correspondingly, the soft breaking terms of the B-LSSM are generally given as
\begin{eqnarray}
&&\mathcal{L}_{soft}=\mathcal{L}_{MSSM}+\Big[-M_{BB^{'}}\tilde{\lambda}_{B^{'}} \tilde{\lambda}_{B} -
\frac{1}{2}M_{B^{'}}\tilde{\lambda}_{B^{'}} \tilde{\lambda}_{B^{'}} -B_{\mu^{'}}\tilde{\eta}_1 \tilde{\eta}_2 +T_{\nu}^{ij} H_2 \tilde{\nu}_i^c \tilde{L}_j+T_x^{ij} \tilde{\eta}_1 \tilde{\nu}_i^c \tilde{\nu}_j^c
\nonumber\\
&&\hspace{1.4cm}
+h.c.\Big]-m_{\tilde{\eta}_1}^2 |\tilde{\eta}_1|^2-m_{\tilde{\eta}_2}^2 |\tilde{\eta}_2|^2-m_{\tilde{\nu},ij}^2(\tilde{\nu}_i^c)^* \tilde{\nu}_j^c,
\end{eqnarray}
with $\lambda_{B}, \lambda_{B^{'}}$ denoting the gaugino of $U(1)_Y$ and $U(1)_{(B-L)}$, respectively.
$\mathcal{L}_{MSSM}$ is the soft breaking terms of the MSSM.

The presence of two Abelian groups gives rise to a new effect absent in the MSSM or other SUSY models with just one Abelian gauge group: the gauge kinetic mixing. It results from the invariance principle which allows the Lagrangian to include a mixing term between the strength tensors of gauge fields associated with the $U(1)$ gauge groups, $-\kappa_{_{Y,BL}}A_{_\mu}^{\prime Y}A^{\prime\mu, BL}$. Here, $A_{_\mu}^{\prime Y}, A^{\prime\mu, BL}$ denote the gauge fields associated with the two $U(1)$ gauge groups, $\kappa_{_{Y,BL}}$ is an antisymmetric tensor which includes the mixing of $U(1)_Y$ and $U(1)_{B-L}$ gauge fields. This mixing couples the $B-L$ sector to the MSSM sector, and even if it is set to zero at $M_{GUT}$, it can be induced through renormalization group equations (RGEs) \cite{RGE1,RGE2,RGE3,RGE4,RGE5,RGE6,RGE7}.  In practice, it turns out that it is easier to work with non-canonical covariant derivatives instead of off-diagonal field-strength tensors. However, both approaches are equivalent \cite{R.F}. Hence in the following, we consider covariant derivatives of the form
\begin{eqnarray}
&&D_\mu=\partial_\mu-i\left(\begin{array}{cc}Y,&B-L\end{array}\right)
\left(\begin{array}{cc}g_{_Y},&g_{_{YB}}^{'}\\g_{_{BY}}^{'},&g_{_{B-L}}\end{array}\right)
\left(\begin{array}{c}A_{_\mu}^{\prime Y} \\ A_{_\mu}^{\prime BL}\end{array}\right)\;,
\label{gauge1}
\end{eqnarray}
where $Y, B-L$ corresponding to the hypercharge and $B-L$ charge respectively. As long as the two Abelian gauge groups are unbroken, we still have the freedom to perform a change of the basis
\begin{eqnarray}
&&D_\mu=\partial_\mu-i\left(\begin{array}{cc}Y,&B-L\end{array}\right)
\left(\begin{array}{cc}g_{_Y},&g_{_{YB}}^{'}\\g_{_{BY}}^{'},&g_{_{B-L}}\end{array}\right)R^TR
\left(\begin{array}{c}A_{_\mu}^{\prime Y} \\ A_{_\mu}^{\prime BL}\end{array}\right)\;,
\label{gauge2}
\end{eqnarray}
where $R$ is a $2\times2$ orthogonal matrix. Choosing $R$ in a proper form, one can write the coupling matrix as
\begin{eqnarray}
&&\left(\begin{array}{cc}g_{_Y},&g_{_{YB}}^{'}\\g_{_{BY}}^{'},&g_{_{B-L}}\end{array}\right)
R^T=\left(\begin{array}{cc}g_{_1},&g_{_{YB}}\\0,&g_{_{B}}\end{array}\right)\;,
\label{gauge3}
\end{eqnarray}
where $g_{_{1}}$ corresponds to the measured hypercharge coupling which is modified in
B-LSSM as given along with $g_{_{B}}$ and $g_{_{YB}}$ in Refs.~\cite{BLSSM1}. In addition, we can redefine the $U(1)$ gauge fields
\begin{eqnarray}
&&R\left(\begin{array}{c}A_{_\mu}^{\prime Y} \\ A_{_\mu}^{\prime BL}\end{array}\right)
=\left(\begin{array}{c}A_{_\mu}^{Y} \\ A_{_\mu}^{BL}\end{array}\right)\;.
\label{gauge4}
\end{eqnarray}

The local gauge symmetry $SU(2)_L\otimes U(1)_Y\otimes U(1)_{B-L}$ breaks down to the electromagnetic symmetry $U(1)_{em}$ as the Higgs fields receive vacuum expectation values (VEVs):
\begin{eqnarray}
&&H_1^1=\frac{1}{\sqrt2}(v_1+{\rm Re}H_1^1+i\,{\rm Im}H_1^1),
\qquad\; H_2^2=\frac{1}{\sqrt2}(v_2+{\rm Re}H_2^2+i\,{\rm Im}H_2^2),\nonumber\\
&&\tilde{\eta}_1=\frac{1}{\sqrt2}(u_1+{\rm Re}\tilde{\eta}_1+i\,{\rm Im}\tilde{\eta}_1),
\qquad\;\quad\;\tilde{\eta}_2=\frac{1}{\sqrt2}(u_2+{\rm Re}\tilde{\eta}_2+i\,{\rm Im}\tilde{\eta}_2)\;.
\end{eqnarray}
For convenience, we define $u^2=u_1^2+u_2^2,\; v^2=v_1^2+v_2^2$ and $\tan\beta^{'}=\frac{u_2}{u_1}$ in analogy to the ratio of the MSSM VEVs ($\tan\beta=\frac{v_2}{v_1}$). Meanwhile, the charged and neutral gauge bosons acquire the nonzero masses as
\begin{eqnarray}
&&\qquad\;\quad\;m_W^2=\frac{1}{4}g_{_2}^2v^2,\nonumber\\
&&\qquad\;\quad\;m_{Z,{Z^{'}}}^2=\frac{1}{8}\Big[(g_{_1}^2+g_2^2+g_{_{YB}}^2)v^2+4g_{_B}^2u^2 \nonumber\\
&&\qquad\;\qquad\;\qquad\;\mp\sqrt{(g_{_1}^2+g_{_2}^2+g_{_{YB}}^2)^2v^4
+8(g_{_{YB}}^2-g_{_1}^2-g_{_2}^2)g_{_B}^2v^2u^2+16g_{_B}^4u^4}\,\Big].
\end{eqnarray}
Compared the MSSM, this new gauge boson $Z'$ makes new contribution to the process $B_s^0\rightarrow \mu^+\mu^-$. In addition, the charged Higgs mass can be written as
\begin{eqnarray}
&&\qquad\;\quad\;M_{H^\pm}^2=\frac{4B_\mu(1+\tan\beta^2)+g_2^2 v_1^2\tan\beta(1+\tan\beta^2)}{4\tan\beta}.
\end{eqnarray}

In the basis (${\rm Re}H_1^1$, ${\rm Re}H_2^2$, ${\rm Re}\tilde{\eta}_1$, ${\rm Re}\tilde{\eta}_2$),
the tree level mass squared matrix for Higgs bosons is given by
\begin{eqnarray}
&&M_h^2=u^2\times\nonumber\\
&&\left(\begin{array}{*{20}{c}}
{\frac{1}{4}\frac{g^2 x^2}{1+\tan\beta^2}+n^2\tan\beta}&{-\frac{1}{4}g^2\frac{x^2\tan\beta}{1+\tan^2\beta}}-n^2&
{\frac{1}{2}g_{_B}g_{_{YB}}\frac{x}{T}}&
{-\frac{1}{2}g_{_B}g_{_{YB}}\frac{x\tan\beta'}{T}}\\ [6pt]
{-\frac{1}{4}g^2\frac{ x^2\tan\beta}{1+\tan^2\beta}}-n^2&{\frac{1}{4}\frac{g^2\tan^2\beta x^2}{1+\tan\beta^2}+\frac{n^2}{\tan\beta}}&
{\frac{1}{2}g_{_B}g_{_{YB}}\frac{x\tan\beta}{T}}&{\frac{1}{2}g_{_B}g_{_{YB}}\frac{x\tan\beta\tan\beta'}{T}}\\ [6pt]
{\frac{1}{2}g_{_B}g_{_{YB}}\frac{x}{T}}&{\frac{1}{2}g_{_B}g_{_{YB}}\frac{x\tan\beta}{T}}&{\frac{g_{_B}^2}{1+\tan^2\beta'}+\tan\beta'N^2}&
{-g_{_B}^2\frac{\tan\beta'}{1+\tan^2\beta'}-N^2}\\ [6pt]
{-\frac{1}{2}g_{_B}g_{_{YB}}\frac{x\tan\beta'}{T}}&{\frac{1}{2}g_{_B}g_{_{YB}}\frac{x\tan\beta\tan\beta'}{T}}&
{-g_{_B}^2\frac{\tan\beta'}{1+\tan^2\beta^{'}}-N^2}&{g_{_B}^2\frac{\tan^2\beta'}{1+\tan^2\beta'}+\frac{N^2}{\tan\beta'}}
\end{array}\right),\nonumber\\
\end{eqnarray}
where $g^2=g_{_1}^2+g_{_2}^2+g_{_{YB}}^2$, $T=\sqrt{1+\tan^2\beta}\sqrt{1+\tan^2\beta'}$,
$n^2=\frac{{\rm Re}B\mu}{u^2}$, $N^2=\frac{{\rm Re}B\mu^{'}}{u^2}$, and $x=\frac{v}{u}$, respectively. These new extra singlets $\tilde{\eta}{_{1,2}}$ and the corresponding pseudo-scalar Higgs boson make new contributions to the process $B_s^0\rightarrow \mu^+\mu^-$, with respect to the MSSM.

Including the leading-log radiative corrections from stop and top particles, the mass of the SM-like Higgs boson can be written as~\cite{HiggsC1,HiggsC2,HiggsC3}
\begin{eqnarray}
&&\Delta m_h^2=\frac{3m_t^4}{2\pi v^2}\Big[\Big(\tilde{t}+\frac{1}{2}+\tilde{X}_t\Big)+\frac{1}{16\pi^2}\Big(\frac{3m_t^2}{2v^2}-32\pi\alpha_3\Big)\Big(\tilde{t}^2
+\tilde{X}_t \tilde{t}\Big)\Big],\nonumber\\
&&\tilde{t}=log\frac{M_S^2}{m_t^2},\qquad\;\tilde{X}_t=\frac{2\tilde{A}_t^2}{M_S^2}\Big(1-\frac{\tilde{A}_t^2}{12M_S^2}\Big),\label{higgs corrections}
\end{eqnarray}
where $\alpha_3$ is the strong coupling constant, $M_S=\sqrt{m_{\tilde t_1}m_{\tilde t_2}}$ with $m_{\tilde t_{1,2}}$ denoting the stop masses, $\tilde{A}_t=A_t-\mu \cot\beta$ with $A_t=T_{u,33}$ being the trilinear Higgs stop coupling and $\mu$ denoting the Higgsino mass parameter. Then the SM-like Higgs mass can be written as
\begin{eqnarray}
&&m_h=\sqrt{(m_{h_1}^0)^2+\Delta m_h^2},\label{higgs mass}
\end{eqnarray}
where $m_{h_1}^0$ denotes the lightest tree-level Higgs boson mass.

Meanwhile, due to the gauge kinetic mixing, additional D-terms contribute to the mass matrices of the squarks and sleptons, and up type squarks affect the subsequent analysis. On the basis $(\tilde u_L, \tilde u_R)$, the mass matrix of up type squarks can be written as
\begin{eqnarray}
&&m_{\tilde u}^2=
\left(\begin{array}{cc}m_{uL},&\frac{1}{\sqrt2}(v_2 T_u^\dagger-v_1\mu Y_u^\dagger)\\\frac{1}{\sqrt2}(v_2 T_u-v_1\mu^* Y_u),&m_{uR}\end{array}\right),
\end{eqnarray}
where,
\begin{eqnarray}
&&m_{uL}=m_{\tilde q}^2+\frac{1}{24}\Big[2g_{_B}(g_{_B}+g_{_{YB}})(u _2^2-u_1^2)+g_{_{YB}}(g_{_B}+g_{_{YB}})(v_2^2-v_1^2)\nonumber\\
&&\qquad\;\quad\;+(-3g_{_2}^2+g_{_1}^2)(v_2^2-v_1^2)\Big]+\frac{v_2^2}{2}Y_u^\dagger Y_u,\nonumber\\
&&m_{uR}=m_{\tilde u}^2+\frac{1}{24}\Big[g_{_B}(g_{_B}+4g_{_{YB}})(u _1^2-u_2^2)+g_{_{YB}}(g_{_B}+4g_{_{YB}})(v_1^2-v_2^2)\nonumber\\
&&\qquad\;\quad\;+4g_{_1}^2(v_1^2-v_2^2)\Big]+\frac{v_1^2}{2}Y_u^\dagger Y_u.
\end{eqnarray}
It can be noted that $\tan\beta'$ and new gauge coupling constants $g_{_B}$, $g_{_{YB}}$ in the B-LSSM can affect the mass matrix of up type squarks. Since up type squarks appear in the loops of the processes $\bar B\rightarrow X_s\gamma$ and $B_s^0\rightarrow \mu^+\mu^-$, new definition of them affects the predictions of $Br(\bar B\rightarrow X_s\gamma)$ and $Br(B_s^0\rightarrow \mu^+\mu^-)$ by influencing their masses and the corresponding couplings.


\section{Theoretical calculation on $Br(\bar B\rightarrow X_s\gamma)$ and $Br(B_s^0\rightarrow \mu^+\mu^-)$\label{sec3}}
\indent\indent
The effective Hamilton for the transition $b\rightarrow s$ at hadronic scale can be written as
\begin{eqnarray}
&&H_{eff}=-\frac{4G_F}{\sqrt{2}}V_{ts}^\ast V_{tb}\Big[C_1\mathcal{O}^c_1+C_2\mathcal{O}_2^c+\sum_{i=3}^6\mathcal{O}_i+\sum_{i=7}^{10}(C_i\mathcal{O}_i+C'_i\mathcal{O}'_i)\nonumber\\
&&\qquad\;\quad\;+\sum_{i=S,P}(C_i\mathcal{O}_i+C'_i\mathcal{O}'_i)\Big],
\end{eqnarray}
where $\mathcal{O}_i(i=1, 2,...,10, S, P)$ and $\mathcal{O}'_i(i=7, 8,...,10, S, P)$ are defined as~\cite{O1,O2,Altmannshofer:2008dz,O3,O4,O6}
\begin{eqnarray}
&&{\cal O}_{_1}^u=(\bar{s}_{_L}\gamma_\mu T^au_{_L})(\bar{u}_{_L}\gamma^\mu T^ab_{_L})\;,\;\;
{\cal O}_{_2}^u=(\bar{s}_{_L}\gamma_\mu u_{_L})(\bar{u}_{_L}\gamma^\mu b_{_L})\;,
\nonumber\\
&&{\cal O}_{_3}=(\bar{s}_{_L}\gamma_\mu b_{_L})\sum\limits_q(\bar{q}\gamma^\mu q)\;,\;\;
{\cal O}_{_4}=(\bar{s}_{_L}\gamma_\mu T^ab_{_L})\sum\limits_q(\bar{q}\gamma^\mu T^aq)\;,
\nonumber\\
&&{\cal O}_{_5}=(\bar{s}_{_L}\gamma_\mu\gamma_\nu\gamma_\rho b_{_L})\sum\limits_q(\bar{q}\gamma^\mu
\gamma^\nu\gamma^\rho q)\;,\;\;
{\cal O}_{_6}=(\bar{s}_{_L}\gamma_\mu\gamma_\nu\gamma_\rho T^ab_{_L})\sum\limits_q(\bar{q}\gamma^\mu
\gamma^\nu\gamma^\rho T^aq)\;,
\nonumber\\
&&{\cal O}_{_7}={e\over 16\pi^2}m_{_b}(\bar{s}_{_L}\sigma_{_{\mu\nu}}b_{_R})F^{\mu\nu}\;,\;\;
{\cal O}_{_7}'={e\over 16\pi^2}m_{_b}(\bar{s}_{_R}\sigma_{_{\mu\nu}}b_{_L})F^{\mu\nu}\;,\;\;
\nonumber\\
&&{\cal O}_{_8}={g_{_s}\over 16\pi^2}m_{_b}(\bar{s}_{_L}\sigma_{_{\mu\nu}}T^ab_{_R})G^{a,\mu\nu}\;,\;\;
{\cal O}_{_8}'={g_{_s}\over 16\pi^2}m_{_b}(\bar{s}_{_R}\sigma_{_{\mu\nu}}T^ab_{_L})G^{a,\mu\nu}\;,\;\;
\nonumber\\
&&{\cal O}_{_9}={e^2\over g_{_s}^2}(\bar{s}_{_L}\gamma_\mu b_{_L})\bar{l}\gamma^\mu l\;,\;\;
{\cal O}_{_9}'={e^2\over g_{_s}^2}(\bar{s}_{_R}\gamma_\mu b_{_R})\bar{l}\gamma^\mu l\;,\;\;
\nonumber\\
&&{\cal O}_{_{10}}={e^2\over g_{_s}^2}(\bar{s}_{_L}\gamma_\mu b_{_L})\bar{l}\gamma^\mu\gamma_5 l\;,\;\;
{\cal O}_{_{10}}'={e^2\over g_{_s}^2}(\bar{s}_{_R}\gamma_\mu b_{_R})\bar{l}\gamma^\mu\gamma_5 l\;,\;\;
\nonumber\\
&&{\cal O}_{_S}={e^2\over16\pi^2}m_{_b}(\bar{s}_{_L}b_{_R})\bar{l}l\;,\;\;
{\cal O}_{_S}'={e^2\over16\pi^2}m_{_b}(\bar{s}_{_R}b_{_L})\bar{l}l\;,\;\;\nonumber\\
&&{\cal O}_{_P}={e^2\over16\pi^2}m_{_b}(\bar{s}_{_L}b_{_R})\bar{l}\gamma_5l\;,\;\;
{\cal O}_{_P}'={e^2\over16\pi^2}m_{_b}(\bar{s}_{_R}b_{_L})\bar{l}\gamma_5l\;,
\label{operators}
\end{eqnarray}
where $g_s$ denotes the strong coupling, $F^{\mu\nu}$ are the electromagnetic  field strength tensors, $G^{\mu\nu}$ are the gluon field strength tensors, $T^a\,(a=1,...,8)$ are $SU(3)$ generators.

\subsection{Rare decay $\bar B\rightarrow X_s\gamma$}

\begin{figure}
\setlength{\unitlength}{1mm}
\centering
\includegraphics[width=4in]{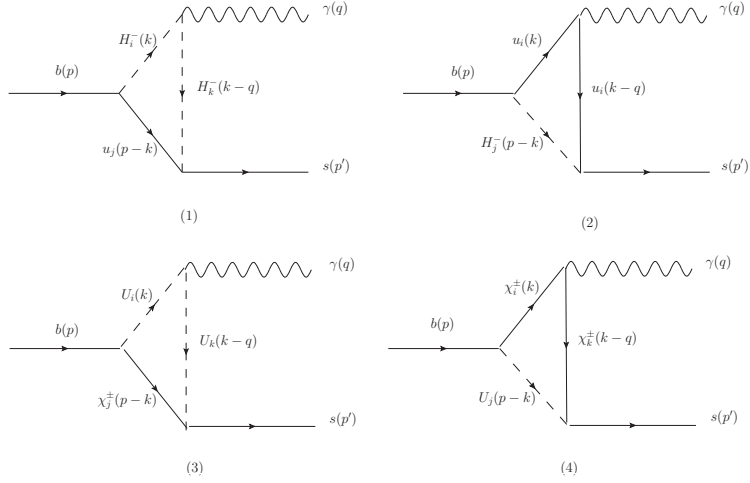}
\vspace{0cm}
\caption[]{The  one loop Feynman diagrams contributing to $\bar{B}\rightarrow X_s\gamma$ from exotic fields in the B-LSSM.}
\label{figBsr}
\end{figure}

\begin{figure}
\setlength{\unitlength}{1mm}
\centering
\includegraphics[width=6in]{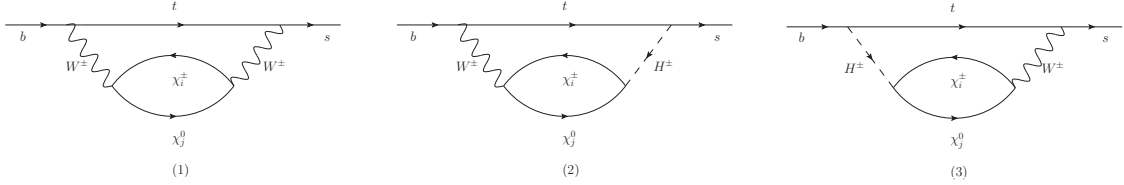}
\vspace{0cm}
\caption[]{The relating two loop diagrams in which a closed heavy fermion loop is attached to virtual $W^\pm$ bosons or $H^\pm$, where a real photon or gluon is attached in all possible ways.}
\label{Bazz diagrams}
\end{figure}

Compared with the SM, the main one loop Feynman diagrams contributing to the process $\bar{B}\rightarrow X_s\gamma$ in the B-LSSM are shown in Fig.~\ref{figBsr}. The picture shows that, new definition of the up type squarks affects the prediction of the process $\bar B\rightarrow X_s\gamma$, with respect to the MSSM. In addition, since the two loop electroweak corrections from closed squark loop are highly suppressed by heavy squark masses, we consider the corrections from closed fermion loop, and the corresponding Feynman diagrams are shown in Fig.~\ref{Bazz diagrams}. From the picture, we can see that, compared with the MSSM, new neutralinos in the B-LSSM make contributions to the process $\bar{B}\rightarrow X_s\gamma$.

Then the branching ratio of $\bar{B}\rightarrow X_s\gamma$ in the B-LSSM can be written as
\begin{eqnarray}
&&Br(\bar{B}\rightarrow X_s\gamma)=R\Big(|C_{7\gamma}(\mu_b)|^2
+N(E_\gamma)\Big)\;,
\end{eqnarray}
where the overall factor $R=2.47\times10^{-3}$, and the nonperturbative contribution
$N(E_\gamma)=(3.6\pm0.6)\times10^{-3}$\cite{H5}. $C_{7\gamma}(\mu_b)$ is defined by
\begin{eqnarray}
&&C_{7\gamma}(\mu_b)=C_{7\gamma,SM}(\mu_b)+C_{7,NP}(\mu_b),
\end{eqnarray}
where we choose the hadron scale $\mu_b=2.5$ GeV and use the SM contribution at NNLO level $C_{7\gamma,SM}(\mu_b) = -0.3689$~\cite{H5,H6,H7,H8}. The Wilson coefficients for new physics at the bottom quark scale can be written as~\cite{H9,H10}
\begin{eqnarray}
&&C_{7,NP}(\mu_b)\approx0.5696
C_{7,NP}(\mu_{EW})+0.1107 C_{8,NP}(\mu_{EW}),
\end{eqnarray}
where
\begin{eqnarray}
&&C_{7,NP}^{NP}(\mu_{EW})=C_{7,NP}^{(1)}(\mu_{EW})+C_{7,NP}^{(2)}(\mu_{EW})+
C_{7,NP}^{(3)}(\mu_{EW})+C_{7,NP}^{(4)}(\mu_{EW})+\nonumber\\
&&\qquad\;\qquad\;\qquad\;C_{7,NP}^{\prime(1)}(\mu_{EW})+C_{7,NP}^{\prime(2)}(\mu_{EW})+C_{7,NP}^{\prime
(3)}(\mu_{EW})+C_{7,NP}^{\prime(4)}(\mu_{EW})+\nonumber\\
&&\qquad\;\qquad\;\qquad\;C_{7,NP}^{WW}(\mu_{EW})+C_{7,NP}^{WH}(\mu_{EW}),\nonumber\\
&&C_{8,NP}(\mu_{EW})=C_{8g,NP}(\mu_{EW})+C_{8g,NP}^{\prime}(\mu_{EW})+C_{8,NP}^{WW}(\mu_{EW})+C_{8,NP}^{WH}(\mu_{EW}),
\end{eqnarray}
 where $C_{7,NP}^{(1,..,4)}(\mu_{EW}),\;C_{7,NP}^{WW}(\mu_{EW}),\;C_{7,NP}^{WH}(\mu_{EW})$ are Wilson coefficients of the process $b\rightarrow s\gamma$ corresponding to Fig.~\ref{figBsr}, Fig.~\ref{Bazz diagrams}, and the concrete expressions are collected in Appendix~\ref{wilsonbsr}. Eq.(\ref{Bazzbsr}) indicates that the corrections to the Wilson coefficients from Fig.~\ref{Bazz diagrams}(2, 3) are suppressed. In addition, $C_{8g,NP}(\mu_{EW}),\;C_{8g,NP}^{\prime}(\mu_{EW}),\;C_{8,NP}^{WW}(\mu_{EW}),\;C_{8,NP}^{WH}(\mu_{EW})$ are Wilson coefficients of the process $b\rightarrow sg$ in the B-LSSM. Compared with the SM, the corresponding Feynman diagrams are shown in Fig.~\ref{figBsr}(2, 3), Fig.~\ref{Bazz diagrams}(1, 2, 3), then $C_{8g,NP}(\mu_{EW})$ and $C_{8g,NP}^{\prime}(\mu_{EW})$ at electroweak scale are
 \begin{eqnarray}
&&C_{8g,NP}(\mu_{EW})=[C_{7,NP}^{(2)}(\mu_{EW})+C_{7,NP}^{(3)}(\mu_{EW})]/Q_u+C_{8,NP}^{WW}(\mu_{EW})+
C_{8,NP}^{WH}(\mu_{EW}), \nonumber\\
&&C_{8g,NP}^{\prime}(\mu_{EW})=C_{8g,NP}(\mu_{EW})(L\leftrightarrow R),
\end{eqnarray}
where $Q_u=2/3$. And the concrete expressions of $C_{8,NP}^{WW}(\mu_{EW}),\;C_{8,NP}^{WH}(\mu_{EW})$ are collected in Appendix~\ref{wilsonbsr}. Compared with the MSSM, new neutralinos make contributions to the process $\bar B\rightarrow X_s\gamma$. In addition, $\tan\beta',\;g_{_B}$ and gauge kinetic mixing $g_{_{YB}}$ in the B-LSSM might influence the theoretical prediction on $Br(\bar B\rightarrow X_s\gamma)$ through affecting the up type squarks masses and the corresponding couplings.

\subsection{Rare decay $B_s^0\rightarrow \mu^+\mu^-$}

\begin{figure}
\setlength{\unitlength}{1mm}
\centering
\includegraphics[width=5.5in]{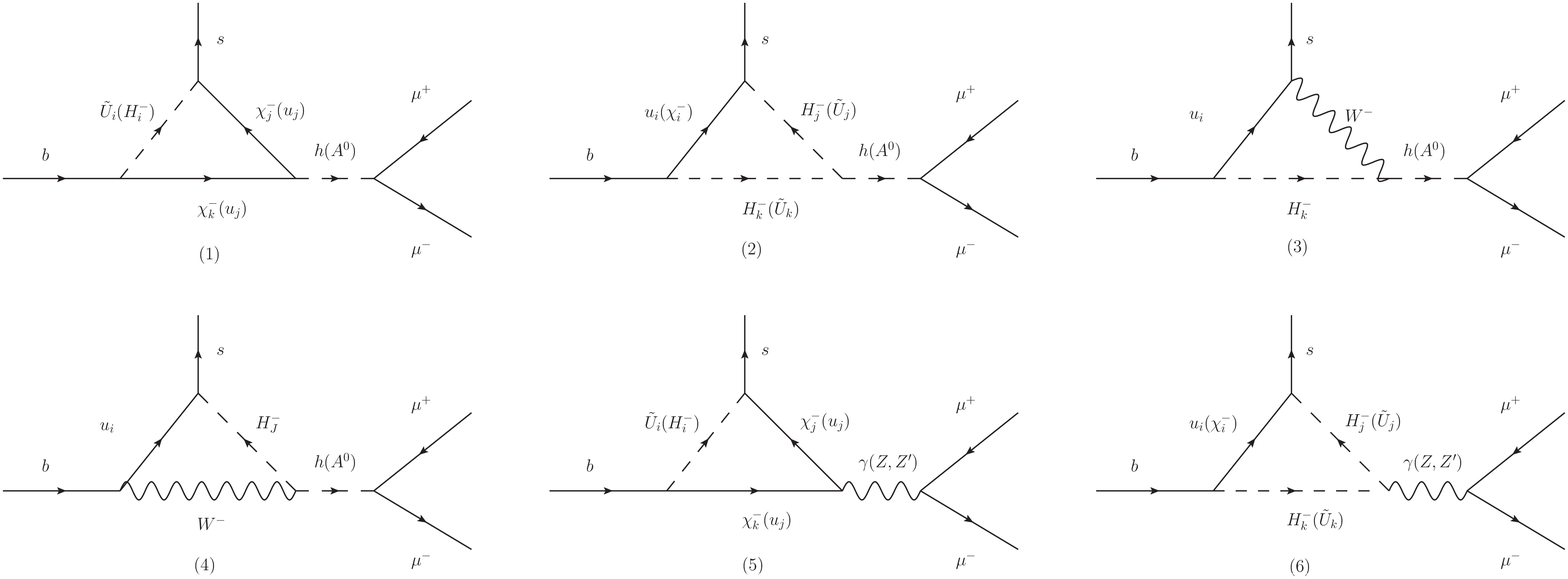}
\vspace{0cm}
\par
\hspace{-0.in}
\includegraphics[width=5.5in]{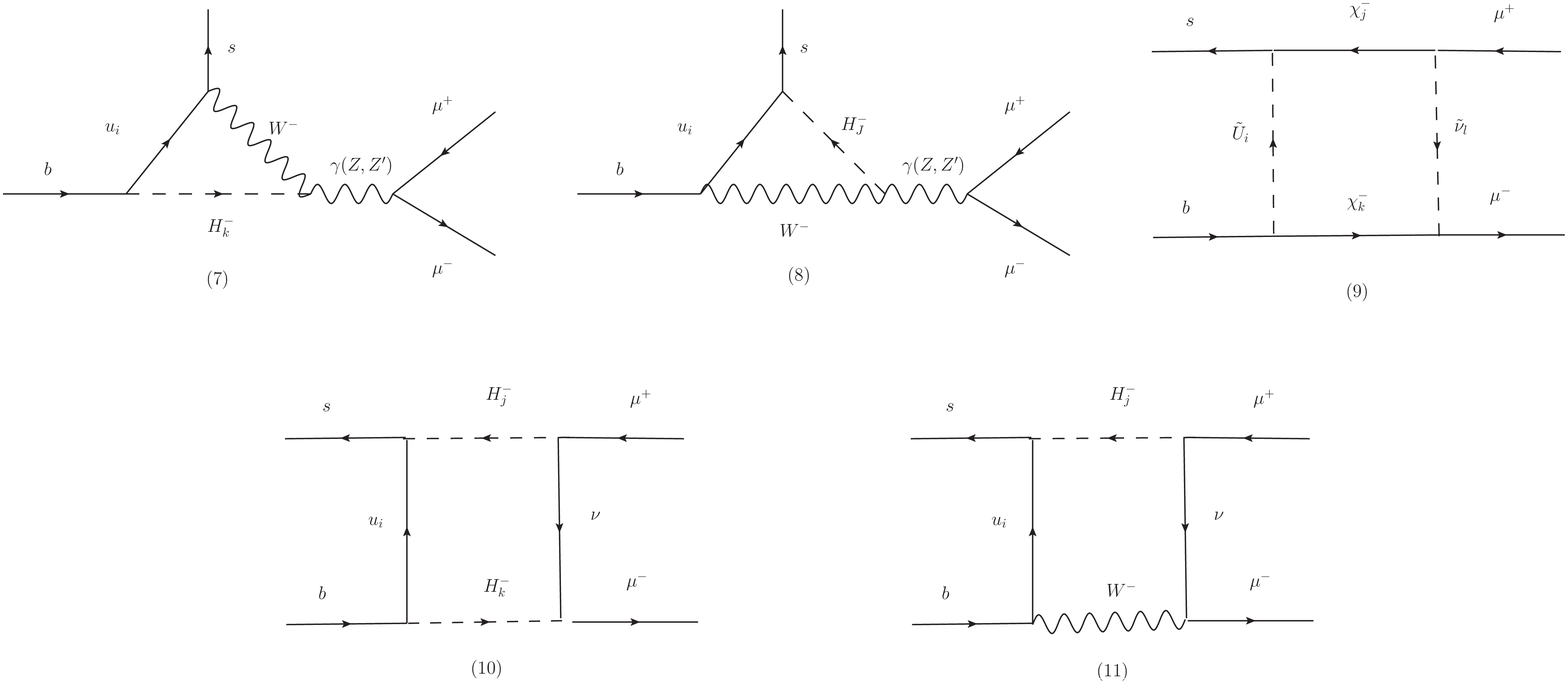}%
\vspace{0cm}
\caption[]{The Feynman diagrams contributing to the decay $B_s^0\rightarrow\mu^+\mu^-$ in the B-LSSM} \label{Bmumu}
\end{figure}

The main Feynman diagrams contributing to the process $B_s^0\rightarrow \mu^+\mu^-$ in the B-LSSM are shown in Fig.~\ref{Bmumu}. Compared with the MSSM, Fig.~\ref{Bmumu} shows that new definition of up type squarks, new gauge boson $Z'$, new scalar and pseudo-scalar Higgs bosons make new contributions to the processes $B_s^0\rightarrow \mu^+\mu^-$ in the B-LSSM.
In addition, since the contributions from Fig.~\ref{Bazz diagrams}(2, 3) are suppressed, we also consider the two loop electroweak corrections from Fig.~\ref{Bazz diagrams}(1), a virtual photon, $Z$ boson or $Z'$ boson is attached in all possible ways. At the electroweak energy scale $\mu_{EW}$, the corresponding Wilson coefficients can be written as
\begin{eqnarray}
&&C_{_{S,NP}}(\mu_{_{\rm EW}})=\frac{\sqrt{2}s_{_W}c_{_W}}{4m_be^3V_{ts}^*V_{tb}}\Big[C_{_{S,NP}}^{(1)}(\mu_{_{\rm EW}})+C_{_{S,NP}}^{(2)}(\mu_{_{\rm EW}})+C_{_{S,NP}}^{(3)}(\mu_{_{\rm EW}})+C_{_{S,NP}}^{(4)}(\mu_{_{\rm EW}})\nonumber\\
&&\qquad\;\qquad\;\qquad\;+C_{_{S,NP}}^{(6)}(\mu_{_{\rm EW}})+C_{_{S,NP}}^{(9)}(\mu_{_{\rm EW}})+C_{_{S,NP}}^{(11)}(\mu_{_{\rm EW}})\Big],\nonumber\\
&&C_{_{S,NP}}^\prime(\mu_{_{\rm EW}})=C_{_{S,NP}}(\mu_{_{\rm EW}})(L\leftrightarrow R),\nonumber\\
&&C_{_{P,NP}}(\mu_{_{\rm EW}})=\frac{\sqrt{2}s_{_W}c_{_W}}{4m_be^3V_{ts}^*V_{tb}}\Big[C_{_{P,NP}}^{(1)}(\mu_{_{\rm EW}})+C_{_{P,NP}}^{(2)}(\mu_{_{\rm EW}})+C_{_{P,NP}}^{(3)}(\mu_{_{\rm EW}})+C_{_{P,NP}}^{(4)}(\mu_{_{\rm EW}})\nonumber\\
&&\qquad\;\qquad\;\qquad\;+C_{_{P,NP}}^{(6)}(\mu_{_{\rm EW}})+C_{_{P,NP}}^{(9)}(\mu_{_{\rm EW}})+C_{_{P,NP}}^{(11)}(\mu_{_{\rm EW}})\Big],\nonumber\\
&&C_{_{P,NP}}^\prime(\mu_{_{\rm EW}})=-C_{_{P,NP}}(\mu_{_{\rm EW}})(L\leftrightarrow R),\nonumber\\
&&C_{_{9,NP}}(\mu_{_{\rm EW}})=\frac{\sqrt{2}s_{_W}c_{_W}g_{_s}^2}{64\pi^2e^3V_{ts}^*V_{tb}}\Big[C_{_{9,NP}}^{(5)}(\mu_{_{\rm EW}})+C_{_{9,NP}}^{(6)}(\mu_{_{\rm EW}})+C_{_{9,NP}}^{(7)}(\mu_{_{\rm EW}})+C_{_{9,NP}}^{(8)}(\mu_{_{\rm EW}})\nonumber\\
&&\qquad\;\qquad\;\qquad\;+C_{_{9,NP}}^{(9)}(\mu_{_{\rm EW}})+C_{_{9,NP}}^{(10)}(\mu_{_{\rm EW}})+C_{_{9,NP}}^{WW}(\mu_{_{\rm EW}})\Big]\;,\nonumber\\
&&C_{_{9,NP}}^\prime(\mu_{_{\rm EW}})=C_{_{9,NP}}(\mu_{_{\rm EW}})(L\leftrightarrow R),\nonumber\\
&&C_{_{10,NP}}(\mu_{_{\rm EW}})=\frac{\sqrt{2}s_{_W}c_{_W}g_{_s}^2}{64\pi^2e^3V_{ts}^*V_{tb}}\Big[C_{_{10,NP}}^{(5)}(\mu_{_{\rm EW}})+C_{_{10,NP}}^{(6)}(\mu_{_{\rm EW}})+C_{_{10,NP}}^{(7)}(\mu_{_{\rm EW}})+C_{_{10,NP}}^{(8)}(\mu_{_{\rm EW}})\nonumber\\
&&\qquad\;\qquad\;\qquad\;+C_{_{10,NP}}^{(9)}(\mu_{_{\rm EW}})+C_{_{10,NP}}^{(10)}(\mu_{_{\rm EW}})+C_{_{10,NP}}^{WW}(\mu_{_{\rm EW}})\Big]\;,\nonumber\\
&&C_{_{10,NP}}^\prime(\mu_{_{\rm EW}})=-C_{_{10,NP}}(\mu_{_{\rm EW}})(L\leftrightarrow R).
\label{Wilson-Coefficients1}
\end{eqnarray}
Here, the superscripts $(1,...,11,WW)$ corresponding to the new physics corrections in Fig.~\ref{Bmumu} and Fig.~\ref{Bazz diagrams}(1), and the concrete expressions of these Wilson coefficients can be found in Appendix~\ref{wilsonbmumu}. The Wilson coefficients at hadronic energy scale from the SM to next-to-next-to-logarithmic (NNLL) accuracy are shown in Table I.
\begin{table}
\begin{tabular}{|c|c|c|c|}
\hline
\hline
$C_{_7}^{eff,SM}$    & $C_{_8}^{eff,SM}$    & $C_{_9}^{eff,SM}$ & $C_{_{10}}^{eff,SM}$\\
\hline
$-0.304$   & $-0.167$  & $4.211$ & $-4.103$\\
\hline
\hline
\end{tabular}
\caption{At hadronic scale $\mu=m_{_b}\simeq4.65$GeV, SM Wilson coefficients to NNLL accuracy. \label{tab1}}
\end{table}
In addition, The Wilson coefficients in Eq.(\ref{Wilson-Coefficients1}) are calculated at the matching scale $\mu_{EW}$, then evolved down to hadronic scale $\mu\sim m_b$ by the renormalization group equations:
\begin{eqnarray}
&&\overrightarrow{C}_{_{NP}}(\mu)=\widehat{U}(\mu,\mu_0)\overrightarrow{C}_{_{NP}}(\mu_0)
\;,\nonumber\\
&&\overrightarrow{C^\prime}_{_{NP}}(\mu)=\widehat{U^\prime}(\mu,\mu_0)
\overrightarrow{C^\prime}_{_{NP}}(\mu_0)
\label{evaluation1}
\end{eqnarray}
with
\begin{eqnarray}
&&\overrightarrow{C}_{_{NP}}^{T}=\Big(C_{_{1,NP}},\;\cdots,\;C_{_{6,NP}},
C_{_{7,NP}}^{eff},\;C_{_{8,NP}}^{eff},\;C_{_{9,NP}}^{eff}-Y(q^2),\;
C_{_{10,NP}}^{eff}\Big)
\;,\nonumber\\
&&\overrightarrow{C}_{_{NP}}^{\prime,\;T}=\Big(C_{_{7,NP}}^{\prime,\;eff},\;
C_{_{8,NP}}^{\prime,\;eff},\;C_{_{9,NP}}^{\prime,\;eff},\;
C_{_{10,NP}}^{\prime,\;eff}\Big)\;.
\label{evaluation2}
\end{eqnarray}
Correspondingly, the evolving matrices are approached as
\begin{eqnarray}
&&\widehat{U}(\mu,\mu_0)\simeq1-\Big[{1\over2\beta_0}\ln{\alpha_{_s}(\mu)\over
\alpha_{_s}(\mu_0)}\Big]\widehat{\gamma}^{(0)T}
\;,\nonumber\\
&&\widehat{U^\prime}(\mu,\mu_0)\simeq1-\Big[{1\over2\beta_0}\ln{\alpha_{_s}(\mu)\over
\alpha_{_s}(\mu_0)}\Big]\widehat{\gamma^\prime}^{(0)T}\;,
\label{evaluation3}
\end{eqnarray}
where the anomalous dimension matrices can be read from Ref. \cite{Gambino1} as
\begin{eqnarray}
&&\widehat{\gamma}^{(0)}=\left(\begin{array}{cccccccccc}
-4&{8\over3}&0&-{2\over9}&0&0&-{208\over243}&{173\over162}&-{2272\over729}&0\\
12&0&0&{4\over3}&0&0&{416\over81}&{70\over27}&{1952\over243}&0\\
0&0&0&-{52\over3}&0&2&-{176\over81}&{14\over27}&-{6752\over243}&0\\
0&0&-{40\over9}&-{100\over9}&{4\over9}&{5\over6}&-{152\over243}&-{587\over162}&-{2192\over729}&0\\
0&0&0&-{256\over3}&0&20&-{6272\over81}&{6596\over27}&-{84032\over243}&0\\
0&0&-{256\over9}&{56\over9}&{40\over9}&-{2\over3}&{4624\over243}&{4772\over81}&-{37856\over729}&0\\
0&0&0&0&0&0&{32\over3}&0&0&0\\
0&0&0&0&0&0&-{32\over9}&{28\over3}&0&0\\
0&0&0&0&0&0&0&0&0&0\\
0&0&0&0&0&0&0&0&0&0\\
\end{array}\right)
\;,\nonumber\\
&&\widehat{\gamma^\prime}^{(0)}=\left(\begin{array}{cccc}
{32\over3}&0&0&0\\
-{32\over9}&{28\over3}&0&0\\
0&0&0&0\\0&0&0&0\\
\end{array}\right)\;.
\label{ADM1}
\end{eqnarray}

Then, the squared amplitude can be written as
\begin{eqnarray}
&&|\mathcal{M}_s|^2=16G_F^2|V_{tb}V_{ts}^*|^2M_{B_s^0}^2\Big[|F_S^s|^2+|F_P^s+2m_{\mu}F_A^s|^2\Big],
\end{eqnarray}
and
\begin{eqnarray}
&&F_S^s=\frac{\alpha_{EW}(\mu_b)}{8\pi}\frac{m_b M_{B_s^0}^2}{m_b+m_s}f_{B_s^0}(C_S-C_S'),\\
&&F_P^s=\frac{\alpha_{EW}(\mu_b)}{8\pi}\frac{m_b M_{B_s^0}^2}{m_b+m_s}f_{B_s^0}(C_P-C_P'),\\
&&F_A^s=\frac{\alpha_{EW}(\mu_b)}{8\pi}f_{B_s^0}\Big[C_{10}^{eff}(\mu_b)-C_{10}^{\prime eff}(\mu_b)\Big],
\end{eqnarray}
where $f_{B_s^0}=(227\pm8){\rm MeV}$ denote the decay constants, $M_{B_s^0}=5.367 \rm GeV$ denote the masses of neutral meson $B_s^0$.

The branching ratio of $B_s^0\rightarrow\mu^+\mu^-$ can be written as
\begin{eqnarray}
&&Br(B_s^0\rightarrow\mu^+\mu^-)=\frac{\tau_{B_s^0}}{16\pi}\frac{|\mathcal{M}_s|^2}{M_{B_s^0}}\sqrt{1-\frac{4m_{\mu}^2}{M_{B_s^0}^2}},
\end{eqnarray}
with $\tau_{B_s^0}=1.466(31){\rm ps}$ denoting the life time of meson.

\section{Numerical analyses\label{sec4}}
\indent\indent
In this section, we present the numerical results of the branching ratios of
rare B-decays $\bar B\rightarrow X_s\gamma$ and $B_s^0\rightarrow \mu^+\mu^-$. The relevant SM input parameters are
chosen as $m_W=80.385{\rm GeV},\;m_Z=90.19{\rm GeV},\;\alpha_{em}(m_Z)=1/128.9,\;\alpha_s(m_Z)=0.118,
\;m_b=4.65{\rm GeV},\;m_s=0.095{\rm GeV}$. Since the tiny neutrino masses basically do not affect $Br(\bar B\rightarrow X_s\gamma)$ and $Br(B_s^0\rightarrow \mu^+\mu^-)$, we take $Y_\nu=Y_x=0$ approximately. The SM-like Higgs mass is \cite{17}
\begin{eqnarray}
&&m_h=125.09\pm0.24{\rm GeV}.
\end{eqnarray}
Meanwhile the CKM matrix is
\begin{eqnarray}
&&\left(\begin{array}{*{20}{c}}
{0.97417}&{0.2248}&{4.09\times10^{-3}}\\ [6pt]
{-0.22}&{0.995}&{4.05\times10^{-2}}\\ [6pt]
{8.2\times10^{-3}}&{-4\times10^{-2}}&{1.009}
\end{array}\right).
\end{eqnarray}

The updated experimental data~\cite{newZ} on searching $Z^\prime$ indicates $M_{Z^{'}}\geq4.05{\rm TeV}$ at 95\% Confidence Level (CL). Due to the contributions of heavy $Z'$ boson are highly suppressed, we choose $M_{Z'}=4.2{\rm TeV}$ in our following numerical analysis. And Refs.~\cite{20,21} give us an upper bound on the ratio between the $Z^{'}$ mass and its gauge coupling at 99\% CL as
\begin{eqnarray}
&&M_{Z^{'}}/g_{_B}\geq6{\rm TeV}\;.
\end{eqnarray}
Then the scope of $g_{_B}$ is limited to $0<g_{_B}\leq0.7$. Additionally, the LHC experimental data also constrain $\tan\beta^{'}<1.5$~\cite{8}. Considering the constraints from the experiments~\cite{17}, for those parameters in Higgsino and gaugino sectors, we appropriately fix $M_{1}=500{\rm GeV},\;M_{2}=600{\rm GeV},\;\mu=700{\rm GeV},\;\mu^{'}=800{\rm GeV},\;M_{BB^{'}}=500{\rm GeV},\;M_{BL}=600{\rm GeV}$, for simplify. For those parameters in the soft breaking terms, we set  $B_\mu'=5\times10^5 {\rm GeV}^2$, $m_{\tilde l}=m_{\tilde e}=T_{\nu}=T_{x}=diag(1, 1, 1) {\rm TeV}$. In addition, the first two generations of squarks are strongly constrained by direct searches at the LHC \cite{ATLAS.PRD,CMS.JHEP} and the third generation squark masses are not constrained by the LHC as strong as the first two generations, and affect the SM-like Higgs mass. Therefore we take $m_{\tilde{q}}=m_{\tilde{u}}=diag(2, 2, m_{\tilde t}){\rm TeV}$, and the discission about the observed Higgs signal in Ref. \cite{add1} limits $m_{\tilde t}\gtrsim1.5{\rm TeV}$. For simplify, we also choose $T_{u_{1,2}}=1{\rm TeV}$. As a key parameter, $T_{u_3}=A_t$ affects SM-like Higgs mass and the following numerical calculation.

\begin{figure}
\setlength{\unitlength}{1mm}
\centering
\includegraphics[width=3.1in]{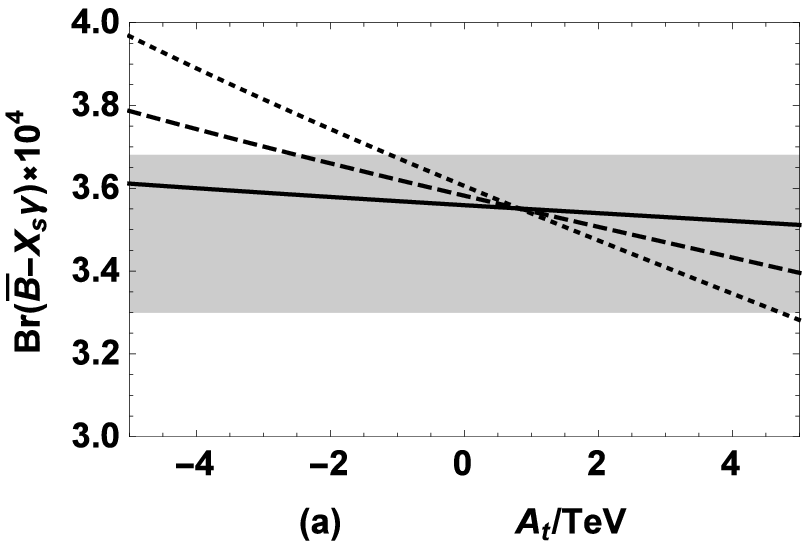}%
\vspace{0.5cm}
\includegraphics[width=3.1in]{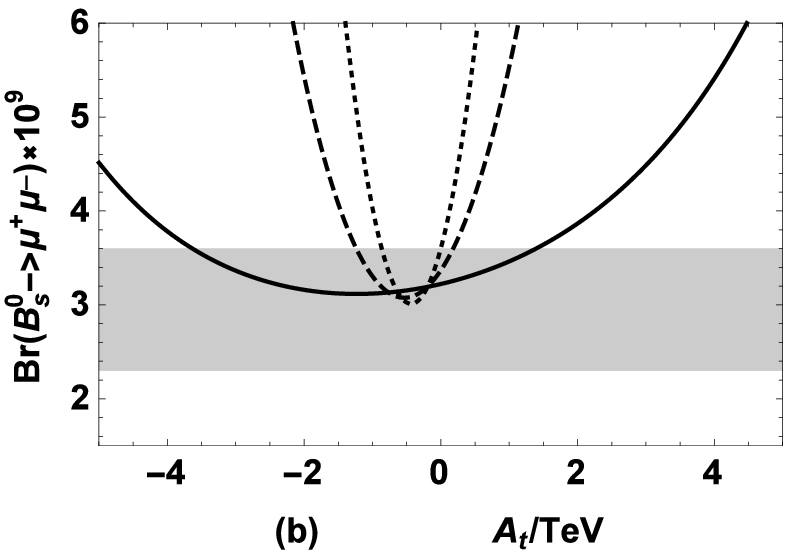}
\vspace{0cm}
\caption[]{$Br(\bar B\rightarrow X_s\gamma)$(a) and $Br(B_s^0\rightarrow\mu^+\mu^-)$(b) versus $A_t$ for $\tan\beta=5({\rm solid \;line}),\;\tan\beta=20({\rm dashed \;line}),\;\tan\beta=35({\rm dotted \;line})$, when $m_{\tilde t}=1.6{\rm TeV},\;M_{H^\pm}=1.5{\rm TeV}$. The gray area denotes the experimental 1$\sigma$ interval.}
\label{figAt}
\end{figure}

In the scenarios of the MSSM, the new physics contributions to the branching ratios of $\bar B\rightarrow X_s\gamma$ and $B_s^0\rightarrow\mu^+\mu^-$ depend essentially on $A_t$, $\tan\beta$ and charged Higgs mass $M_{H^\pm}$. In order to see how $A_t$, $\tan\beta$, $M_{H^\pm}$ affect the theoretical evaluations of $Br(\bar B\rightarrow X_s\gamma)$ and $Br(B_s^0\rightarrow\mu^+\mu^-)$ in the B-LSSM, we assume that $g_{_B}=0.4,\;g_{_{YB}}=-0.5,\;\tan\beta'=1.1$ in the following analysis. Then taking $m_{\tilde t}=1.6{\rm TeV},\;M_{H^\pm}=1.5{\rm TeV}$, we plot $Br(\bar B\rightarrow X_s\gamma)$ and $Br(B_s^0\rightarrow\mu^+\mu^-)$ versus $A_t$ in Fig.~\ref{figAt}, for $\tan\beta=5$(solid line), $\tan\beta=20$(dashed line) and $\tan\beta=35$(dotted line), respectively. The gray area denotes the experimental 1$\sigma$ bounds in Eq.(\ref{experimental data}). In Fig.~\ref{figAt}(a), we can see that $Br(\bar B\rightarrow X_s\gamma)$ decreases with the increasing of $A_t$, and $Br(\bar B\rightarrow X_s\gamma)$ will be easily coincides with experimental data within one standard deviation when $A_t$ is positive. Meanwhile, Fig.~\ref{figAt}(b) shows that $Br(B_s^0\rightarrow\mu^+\mu^-)$ favor $A_t$ in the ranges $-3.6{\rm TeV}\lesssim A_t\lesssim1.4{\rm TeV}$ as $\tan\beta=5$, $-1.2{\rm TeV}\lesssim A_t\lesssim0.2{\rm TeV}$ as $\tan\beta=20$ and $-0.8{\rm TeV}\lesssim A_t\lesssim0{\rm TeV}$ as $\tan\beta=35$, which also coincide with the experimental data on $Br(\bar B\rightarrow X_s\gamma)$. It can be noted that when $\tan\beta$ is large, the range of $A_t$ is limited strongly by the experimental data on $Br(B_s^0\rightarrow\mu^+\mu^-)$.

\begin{figure}
\setlength{\unitlength}{1mm}
\centering
\includegraphics[width=3.1in]{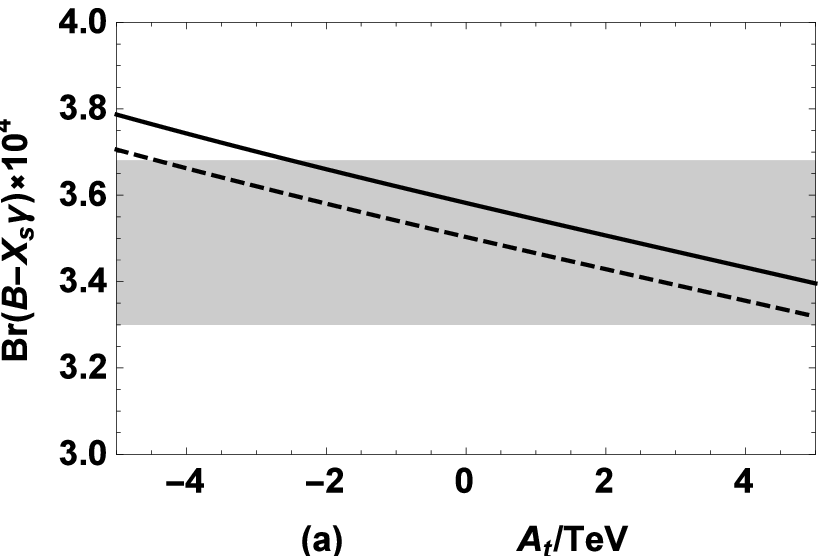}%
\vspace{0.5cm}
\includegraphics[width=3.1in]{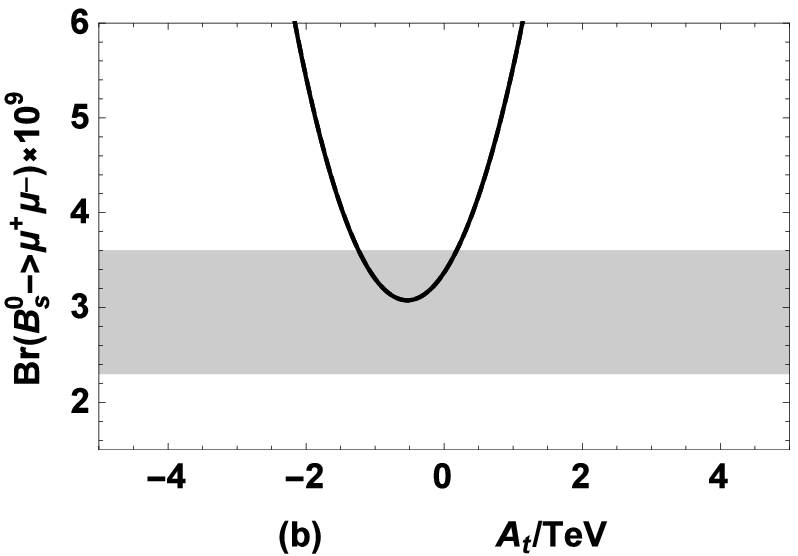}
\vspace{0cm}
\caption[]{$Br(\bar B\rightarrow X_s\gamma)$(a) and $Br(B_s^0\rightarrow\mu^+\mu^-)$(b) versus $A_t$ for two loop result (solid line) and one loop result (dashed line). The gray area denotes the experimental 1$\sigma$ interval.}
\label{figcompareAt}
\end{figure}

Then, in order to see the difference between two loop and one loop corrections to the processes, we take $\tan\beta=20$, and plot $Br(\bar B\rightarrow X_s\gamma)$, $Br(B_s^0\rightarrow\mu^+\mu^-)$ versus $A_t$ in Fig.~\ref{figcompareAt}. The solid and dashed line denote two loop and one loop predictions respectively.
Fig.~\ref{figcompareAt}(a) shows that, the relative corrections from two loop diagrams to one loop corrections of $Br(\bar B\rightarrow X_s\gamma)$ can reach around $3\%$, which produces a more precise prediction on the process $\bar B\rightarrow X_s\gamma$, and we cannot neglect the corrections with this magnitude. In Fig.~\ref{figcompareAt}(b), these two lines coincide with each other, which indicates the two loop corrections to $Br(B_s^0\rightarrow\mu^+\mu^-)$ are negligible compared with the one loop corrections. In the analysis of the numerical results, we use the more precise two loop predictions.

\begin{figure}
\setlength{\unitlength}{1mm}
\centering
\includegraphics[width=3.3in]{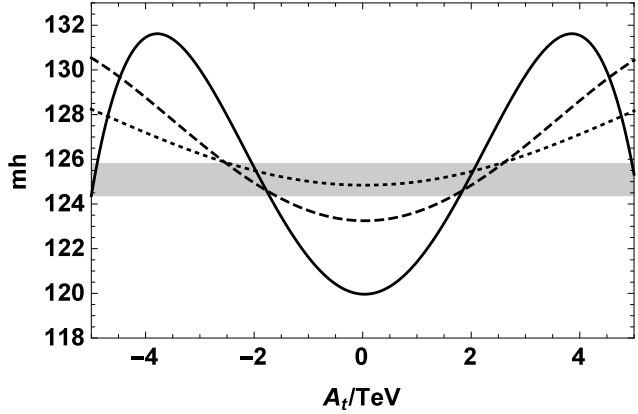}%
\vspace{0.5cm}
\caption[]{Taking $\tan\beta=20$, we plot the SM-like Higgs mass $m_h$ versus $A_t$ for $m_{\tilde t}=1.5{\rm TeV}$(solid line), $m_{\tilde t}=2.5{\rm TeV}$(dashed line) and $m_{\tilde t}=3.5{\rm TeV}$(dotted line), where the gray area denotes the experimental $3\sigma$ interval.}
\label{figmh}
\end{figure}
In addition, we also need to consider the constraint of the SM-like Higgs mass. Taking $\tan\beta=20$, we plot the SM-like Higgs mass $m_h$ versus $A_t$ in Fig.~\ref{figmh} for $m_{\tilde t}=1.5{\rm TeV}$(solid line), $m_{\tilde t}=2.5{\rm TeV}$(dashed line) and $m_{\tilde t}=3.5{\rm TeV}$(dotted line). The gray area denotes the experimental $3\sigma$ interval. To keep the SM-like Higgs mass around $125{\rm GeV}$, we need $A_t\approx\pm1.8{\rm TeV}$ as $m_{\tilde t}=1.5{\rm TeV}$. When $m_{\tilde t}=2.5{\rm TeV}$, we require that $A_t$ in the range $-2.5{\rm TeV}\lesssim A_t\lesssim1.6{\rm TeV}$ or $1.6{\rm TeV}\lesssim A_t\lesssim2.5{\rm TeV}$. And the allowed range of $A_t$ is $-2.5{\rm TeV}\lesssim A_t\lesssim2.5{\rm TeV}$ when $m_{\tilde t}=3.5{\rm TeV}$.

\begin{figure}
\setlength{\unitlength}{1mm}
\centering
\includegraphics[width=3.1in]{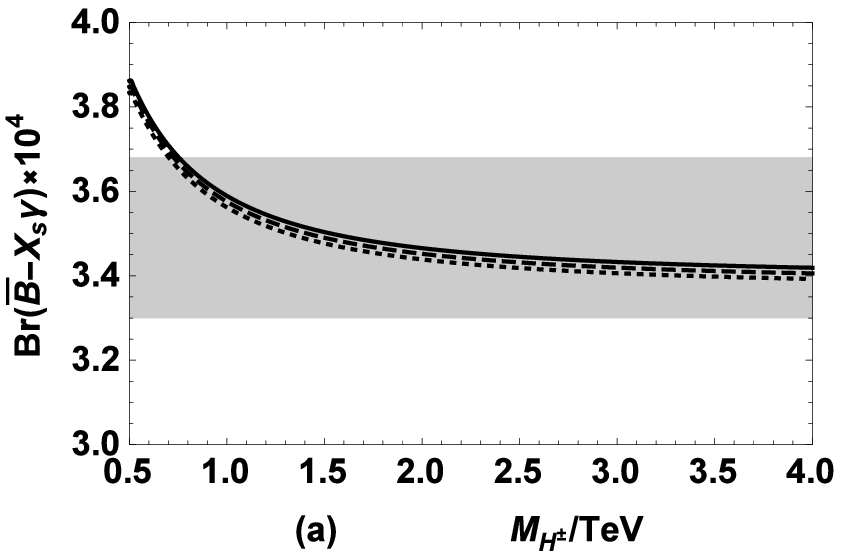}%
\vspace{0.5cm}
\includegraphics[width=3.1in]{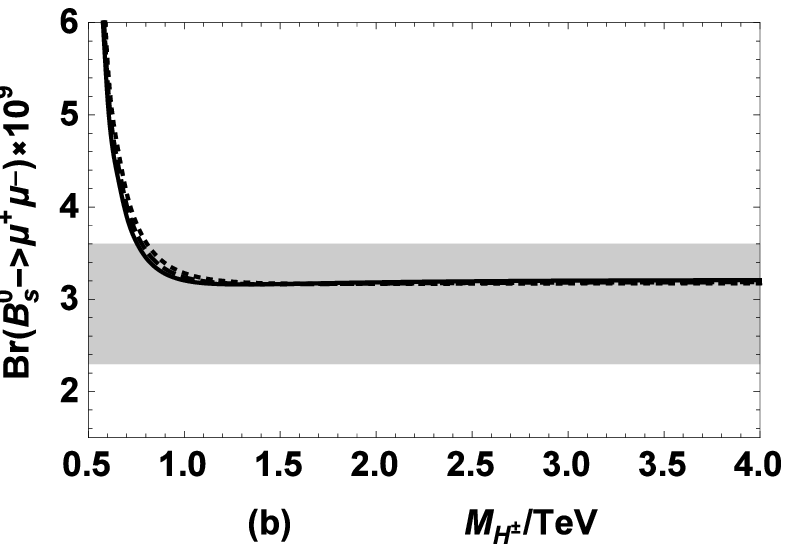}
\vspace{0cm}
\caption[]{$Br(\bar B\rightarrow X_s\gamma)$(a) and $Br(B_s^0\rightarrow\mu^+\mu^-)$(b) versus $M_{H^\pm}$ for $A_t=-2.5{\rm TeV}({\rm solid \;line}),\;A_t=-0.5{\rm TeV}({\rm dashed \;line}),\;A_t=1.5{\rm TeV}({\rm dotted \;line})$, when $\tan\beta=20,\;m_{\tilde t}=3.5{\rm TeV}$. The gray area denotes the experimental 1$\sigma$ interval.}
\label{figmch}
\end{figure}

Since the large charged Higgs mass does not affect the SM-like Higgs mass signally, we can choose $A_t=-2.5{\rm TeV},\;-0.5{\rm TeV}$ and $1.5{\rm TeV}$ for $m_{\tilde t}=3.5{\rm TeV}$, to keep the SM-like Higgs mass around $125{\rm GeV}$. Then we plot $Br(\bar B\rightarrow X_s\gamma)$ and $Br(B_s^0\rightarrow\mu^+\mu^-)$ versus $M_{H^\pm}$ in Fig.~\ref{figmch}, where the solid line, dashed line and dotted line denote $A_t=-2.5{\rm TeV},\;-0.5{\rm TeV},\;1.5{\rm TeV}$, respectively. The gray area denotes the experimental 1$\sigma$ bounds.
It is obvious that $A_t$ affect the numerical results negligibly compared with $M_{H^\pm}$, because large $m_{\tilde t}$ suppresses the effects of $A_t$. In addition, Fig.~\ref{figmch}(a) shows that $Br(\bar B\rightarrow X_s\gamma)$ decreases along with the increasing of $M_{H^\pm}$, because the contributions from charged Higgs diagrams decay like $1/M_{H^\pm}^4$\cite{MH}. And the experimental data on $Br(\bar B\rightarrow X_s\gamma)$ favor $M_{H^\pm}$ lying in the ranges $M_{H^\pm}\gtrsim0.7{\rm TeV}$. Meanwhile, in Fig.~\ref{figmch}(b), we can see that for small $M_{H^\pm}$, the new physics could contribute with large corrections to the branching ratio of $B_s^0\rightarrow\mu^+\mu^-$, and the experimental data on $Br(B_s^0\rightarrow\mu^+\mu^-)$ limits that $M_{H^\pm}\gtrsim0.8{\rm TeV}$.

\begin{figure}
\setlength{\unitlength}{1mm}
\centering
\includegraphics[width=3.1in]{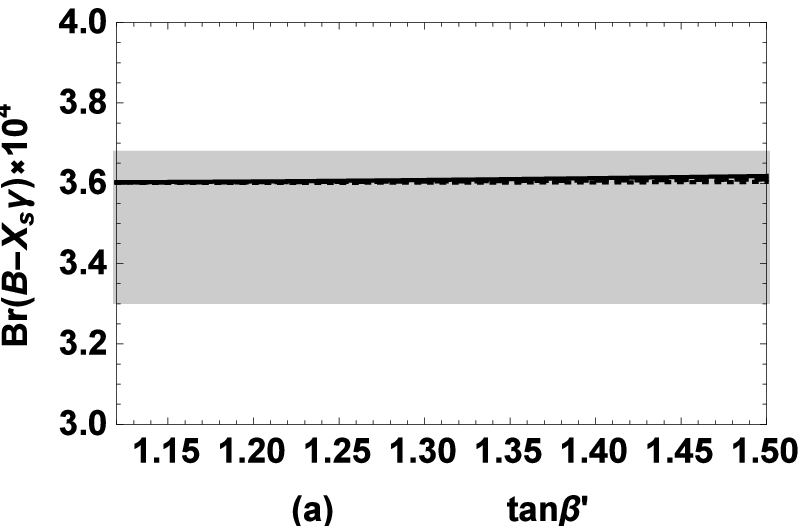}%
\vspace{0.5cm}
\includegraphics[width=3.1in]{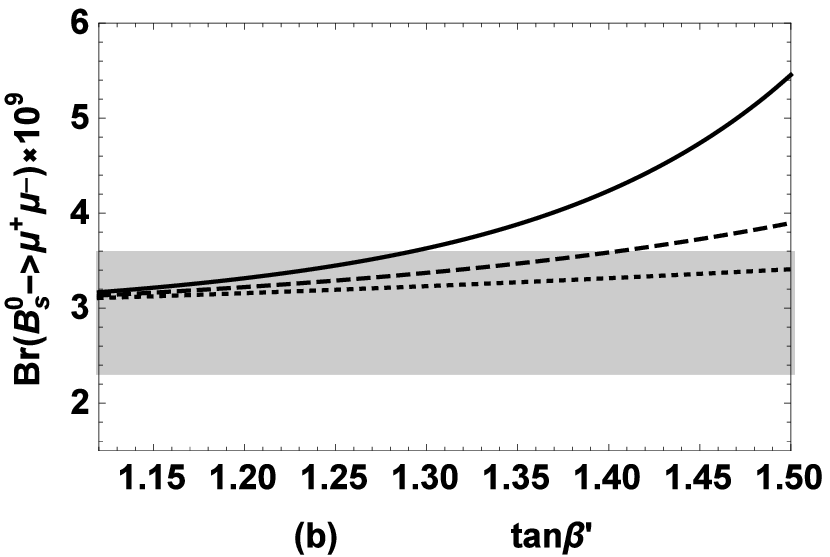}
\vspace{0cm}
\caption[]{$Br(\bar B\rightarrow X_s\gamma)$(a) and $Br(B_s^0\rightarrow\mu^+\mu^-)$(b) versus $\tan\beta'$ for $g_{_{YB}}=-0.6({\rm solid \;line}),\;g_{_{YB}}=-0.5({\rm dashed \;line}),\;g_{_{YB}}=-0.4({\rm dotted \;line})$. The gray area denotes the experimental 1$\sigma$ interval.}
\label{figtB}
\end{figure}

In order to see the effects of $g_{_{B}},\;g_{_{YB}}$ and $\tan\beta'$ which are new parameters in the B-LSSM compared with MSSM, we take $\tan\beta=20,\;m_{\tilde t}=1.6{\rm TeV},\;A_t=-0.5{\rm TeV},\;M_{H^\pm}=1.5{\rm TeV}$ and $g_{_B}=0.2$. Then we plot $Br(\bar B\rightarrow X_s\gamma)$ and $Br(B_s^0\rightarrow\mu^+\mu^-)$ varying with $\tan\beta'$ in Fig.~\ref{figtB}, for $g_{_{YB}}=-0.6({\rm solid \;line}),\;g_{_{YB}}=-0.5({\rm dashed \;line}),\;g_{_{YB}}=-0.4({\rm dotted \;line})$, respectively. Considering the constraint from concrete Higgs boson mass, the allowed range of $\tan\beta'$ is $1.12<\tan\beta'<1.5$. The gray area denotes the experimental 1$\sigma$ bounds.
The picture shows that the numerical results of $Br(\bar B\rightarrow X_s\gamma)$ depend on $g_{_{YB}}$ or $\tan\beta'$ weakly in our chosen parameter space. Meanwhile, $Br(B_s^0\rightarrow\mu^+\mu^-)$ depends on $\tan\beta'$ more acutely when $|g_{_{YB}}|$ is lager, and can exceed the experimental $1\sigma$ upper bound easily with the increasing of $\tan\beta'$ when $g_{_{YB}}=-0.6$, which indicates that the effect of $\tan\beta'$ to the process is influenced by the strength of gauge kinetic mixing strongly. $\tan\beta'$ affects $Br(B_s^0\rightarrow\mu^+\mu^-)$ mainly through the new definition of up type squarks, new scalar and pseudoscalar Higgs bosons. In addition, $Br(B_s^0\rightarrow\mu^+\mu^-)$ increases with the increasing of $|g{_{YB}}|$ and depend on $g{_{YB}}$ acutely when $\tan\beta'$ is large.

\begin{figure}
\setlength{\unitlength}{1mm}
\centering
\includegraphics[width=3.1in]{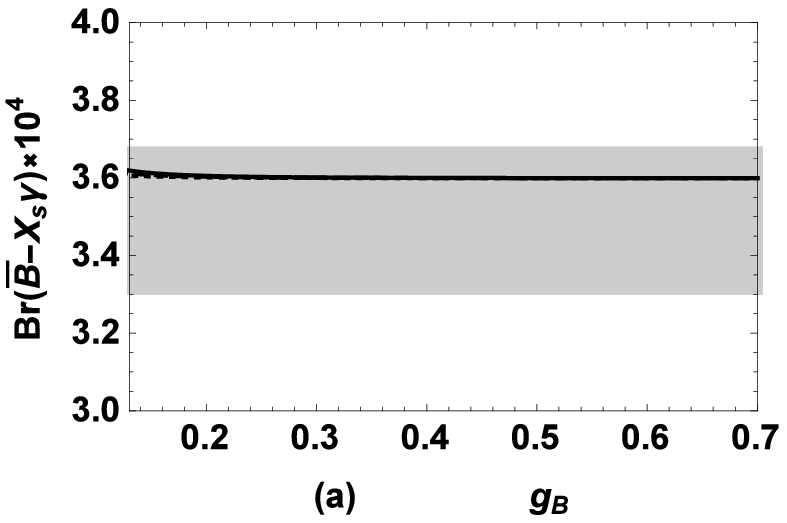}%
\vspace{0.5cm}
\includegraphics[width=3.1in]{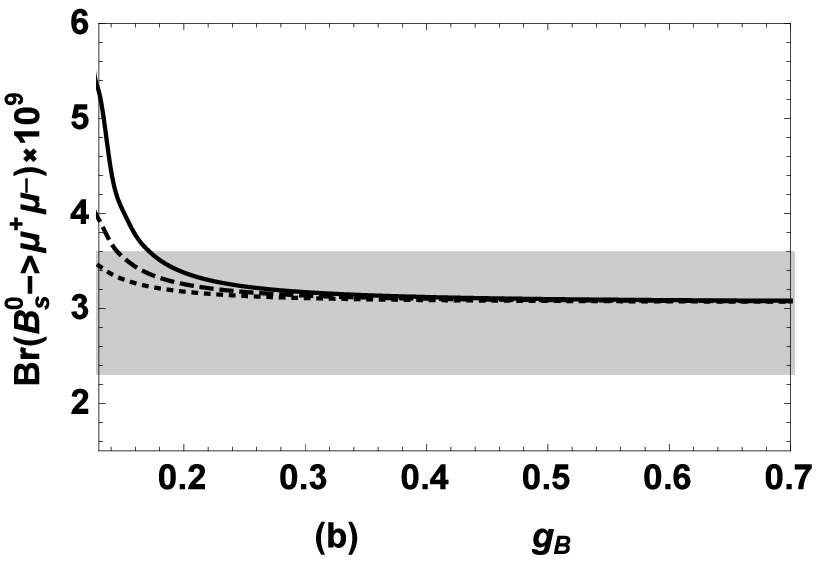}
\vspace{0cm}
\caption[]{$Br(\bar B\rightarrow X_s\gamma)$(a) and $Br(B_s^0\rightarrow\mu^+\mu^-)$(b) versus $g_{_{B}}$ for $g_{_{YB}}=-0.6({\rm solid \;line}),\;g_{_{YB}}=-0.5({\rm dashed \;line}),\;g_{_{YB}}=-0.4({\rm dotted \;line})$. The gray area denotes the experimental 1$\sigma$ interval.}
\label{figgB}
\end{figure}

Then we take $\tan\beta'=1.2$, and plot $Br(\bar B\rightarrow X_s\gamma)$ and $Br(B_s^0\rightarrow\mu^+\mu^-)$ varying with $g_{_{B}}$ in Fig.~\ref{figgB}, for $g_{_{YB}}=-0.6({\rm solid \;line}),\;g_{_{YB}}=-0.5({\rm dashed \;line}),\;g_{_{YB}}=-0.4({\rm dotted \;line})$, respectively. Considering the constraints from the concrete Higgs boson mass, the allowed range of $g_{_B}$ is $0.13<g_{_B}<0.7$. The gray area denotes the experimental 1$\sigma$ bounds.
It can be noted that $g_{_B}$ and $g_{_{YB}}$ do not affect $Br(\bar B\rightarrow X_s\gamma)$ obviously. And $Br(B_s^0\rightarrow\mu^+\mu^-)$ can exceed the experimental $1\sigma$ upper bound easily when $g_{_B}$ is small and $|g_{_{YB}}|$ is large. In addition, with the decreasing of $|g_{_{YB}}|$, $Br(B_s^0\rightarrow\mu^+\mu^-)$ depends on $g_{_B}$ negligibly, which indicates that the effect of $g_{_B}$ to the process is influenced by the strength of gauge kinetic mixing strongly. $g_{_B}$ and $g_{_{YB}}$ affect $Br(B_s^0\rightarrow\mu^+\mu^-)$ mainly in two ways. Firstly,
$g_{_B}$ and $g_{_{YB}}$ affect the up type squark masses and the corresponding rotation matrix, which appears in the couplings involve the up type squarks. Secondly, they affect the process $B_s^0\rightarrow\mu^+\mu^-$ by influencing the new contributions from $Z'$ gauge boson, new scalar and pseudoscalar Higgs bosons in Fig.~\ref{Bmumu}.

\section{Summary\label{sec5}}
\indent\indent
Rare B-meson decays offer high sensitivity to new physics beyond SM. In this work, we study the two loop electroweak corrections to the branching ratios $Br(\bar B\rightarrow X_s\gamma)$ and $Br(B_s^0\rightarrow\mu^+\mu^-)$ in the framework of the B-LSSM under a minimal flavor violating assumption. Considering the constraint from the observed Higgs signal and updated experimental data, the numerical analyses indicate that the corrections from two loop diagrams to the process $\bar B\rightarrow X_s\gamma$ can reach around $3\%$, which produces a more precise prediction on the process $\bar B\rightarrow X_s\gamma$. Nevertheless, the corrections from two loop diagrams to the process $B_s^0\rightarrow\mu^+\mu^-$ are negligible, compared with the one loop corrections. Meanwhile, the new physics can fit the experimental data for the rare decay $\bar B\rightarrow X_s\gamma$ and $B_s^0\rightarrow\mu^+\mu^-$, and also further constrain the parameter space of the model. Under our assumption on parameters of the considered model, $A_t,\;\tan\beta,\;M_{H^\pm}$ affect the theoretical predictions on $Br(\bar B\rightarrow X_s\gamma)$ and $Br(B_s^0\rightarrow\mu^+\mu^-)$ obviously. And when $\tan\beta$ is large, $A_t$ is limited strongly by the experimental data for $Br(B_s^0\rightarrow\mu^+\mu^-)$. In addition, $\tan\beta',\;g_{_{YB}}$ and $g_{_B}$ can also affect theoretical predictions on $Br(B_s^0\rightarrow\mu^+\mu^-)$ obviously.

\begin{acknowledgments}
\indent\indent
The work has been supported by the National Natural Science Foundation of China (NNSFC) with Grants No. 11535002,  No. 11705045, and No. 11647120, Natural Science Foundation of Hebei province with Grants No. A2016201010 and No. A2016201069, Foundation of Department of Education of Liaoning
province with Grant No. 2016TSPY10, Youth
Foundation of the University of Science and Technology
Liaoning with Grant No. 2016QN11, Hebei Key Lab of Optic-Eletronic Information and Materials, and the Midwest Universities Comprehensive Strength Promotion
project.
\end{acknowledgments}

\appendix

\section{The Wilson coefficients of the process $\bar B\rightarrow X_s\gamma$. \label{wilsonbsr}}
The one loop Wilson coefficients corresponding to $b\rightarrow s\gamma$ can be written as
\begin{eqnarray}
&&C_{7,NP}^{(1)}(\mu_{EW})=\sum_{H^-_i,u_j}\frac{s_W^2}{2e^2V^*_{ts}V_{tb}} \Big\{ \frac{1}{2}C_{H^-_i\bar s u_j}^R C_{H^-_i b \bar u_j}^{L}[-I_3(x_{u_j},x_{H^-_i})+I_4(x_{u_j},x_{H^-_i})]+\nonumber\\
&&\qquad\qquad\qquad\quad\frac{m_{u_j}}{m_b}C_{H^-_i\bar s u_j}^L C_{H^-_i b \bar u_j}^{L}[-I_1(x_{u_j},x_{H^-_i})+I_3(x_{u_j},x_{H^-_i})]\Big\},\nonumber\\
&&C_{7,NP}^{(2)}(\mu_{EW})=\sum_{H^-_j,u_i}\frac{s_W^2}{3e^2V^*_{ts}V_{tb}} \Big\{ \frac{1}{2}C_{H^-_j\bar s u_i}^R C_{H^-_j b \bar u_i}^{L}[-I_1(x_{u_i},x_{H^-_j})+2I_3(x_{u_i},x_{H^-_j})\nonumber\\
&&\qquad\qquad\qquad\quad-I_4(x_{u_i},x_{H^-_j})]+\frac{m_{u_i}}{m_b}C_{H^-_j\bar s u_i}^L C_{H^-_j b \bar u_i}^{L}[I_1(x_{u_i},x_{H^-_j})-I_2(x_{u_i},x_{H^-_j})\nonumber\\
&&\qquad\qquad\qquad\quad-I_3(x_{u_i},x_{H^-_j})]\Big\},\nonumber\\
&&C_{7,NP}^{(3)}(\mu_{EW})=\sum_{U^+_i,\chi^-_j}\frac{s_W^2}{3e^2V^*_{ts}V_{tb}} \Big\{ \frac{1}{2}C_{U^+_i\bar s \chi^-_j}^R C_{U^+_i b \bar \chi^-_j}^{L}[-I_3(x_{\chi^-_j},x_{U^+_i})+I_4(x_{\chi^-_j},x_{U^+_i})]+\nonumber\\
&&\qquad\qquad\qquad\quad\frac{m_{\chi^-_j}}{m_b}C_{U^+_i\bar s \chi^-_j}^L C_{U^+_i b \bar \chi^-_j}^{L}[-I_1(x_{\chi^-_j},x_{U^+_i})+I_3(x_{\chi^-_j},x_{U^+_i})]\Big\},\nonumber\\
&&C_{7,NP}^{(4)}(\mu_{EW})=\sum_{U^+_j,\chi^-_i}\frac{s_W^2}{2e^2V^*_{ts}V_{tb}} \Big\{ \frac{1}{2}C_{U^+_i\bar s \chi^-_i}^R C_{U^+_j b \bar \chi^-_i}^{L}[-I_1(x_{\chi^-_i},x_{U^+_j})+2I_2(x_{\chi^-_i},x_{U^+_j})\nonumber\\
&&\qquad\qquad\qquad\quad-I_4(x_{\chi^-_i},x_{U^+_j})]+\frac{m_{\chi^-_i}}{m_b}C_{U^+_j\bar s \chi^-_i}^L C_{U^+_j b \bar \chi^-_i}^{L}[I_1(x_{\chi^-_i},x_{U^+_j})-I_2(x_{\chi^-_i},x_{U^+_j})\nonumber\\
&&\qquad\qquad\qquad\quad-I_3(x_{\chi^-_i},x_{U^+_j})]\Big\},\nonumber\\
&&C_{7}^{\prime NP(a)}(\mu_{EW})=C_{7}^{\prime NP(a)}(\mu_{EW})(L\leftrightarrow R), (a=1,2,3,4),
\end{eqnarray}
where $S$ dentes CP-even and CP-odd Higgs, $C_{abc}^{L,R}$ denotes the constant parts of the interaction vertex about $abc$, which can be got through SARAH, and $a, b, c$ denote the interactional particles. $L$ and $R$ in superscript denote the left-hand part and right-hand part. Denoting $x_i=\frac{m_i^2}{m_W^2}$, the concrete expressions for $I_k(k=1,...,4)$ can be given as:
\begin{eqnarray}
&&I_1(x_1,x_2)=\frac{1+{\rm ln} x_2}{(x_2-x_1)}+\frac{x_1 {\rm ln} x_1-x_2 {\rm ln} x_2}{(x_2-x_1)^2},\nonumber\\
&&I_2(x_1,x_2)=-\frac{1+{\rm ln} x_1}{(x_2-x_1)}-\frac{x_1 {\rm ln} x_1-x_2 {\rm ln} x_2}{(x_2-x_1)^2},\nonumber\\
&&I_3(x_1,x_2)=\frac{1}{2}\Big[\frac{3+2{\rm ln} x_2}{(x_2-x_1)}-\frac{2x_2+4x_2{\rm ln} x_2}{(x_2-x_1)^2}-\frac{2x_1^2{\rm ln} x_1}{(x_2-x_1)^3}+\frac{2x_2^2{\rm ln} x_2}{(x_2-x_1)^3}\Big],\nonumber\\
&&I_4(x_1,x_2)=\frac{1}{4}\Big[\frac{11+6{\rm ln} x_2}{(x_2-x_1)}-\frac{15+18x_2{\rm ln} x_2}{(x_2-x_1)^2}+\frac{6x_2^2+18x_2^2{\rm ln} x_2}{(x_2-x_1)^3}+\nonumber\\
&&\qquad\qquad\quad\frac{6x_1^3{\rm ln} x_1-6x_2^3{\rm ln} x_2}{(x_2-x_1)^4}\Big].
\end{eqnarray}

Assuming $m_{\chi_i^\pm},\;m_{\chi_j^0}\gg m_{_W}$, then the two loop Wilson coefficients corresponding to $b\rightarrow s\gamma$ can be simplified as
\begin{eqnarray}
&&C_{7,NP}^{WW}(\mu_{EW})=\sum_{\chi^\pm_i,\chi^0_j}\frac{-1}{8\pi^2} \Big\{ (|C_{W^-\bar\chi^0_j \chi^+_i}^{L}|^2 +|C_{W^-\bar\chi^0_j \chi^+_i}^{R}|^2)\Big[ P(\frac{1}{8},\frac{1}{4},\frac{-1}{48},\frac{-1}{144},1,x_t)+\nonumber\\
&&\qquad\qquad\qquad\quad \frac{2}{3}P(\frac{-11}{12},\frac{-29}{72},\frac{-1}{12},\frac{1}{144},1,x_t)\Big]+(C_{W^-\bar\chi^0_j \chi^+_i}^{R} C_{W^-\bar\chi^0_j \chi^+_i}^{L*}+C_{W^-\bar\chi^0_j \chi^+_i}^{L}\nonumber\\
&&\qquad\qquad\qquad\quad C_{W^-\bar\chi^0_j \chi^+_i}^{R*})\Big[ P(\frac{1}{16},\frac{1}{4},\frac{-1}{16},\frac{1}{144},1,x_t)+\frac{2}{3}P(\frac{-11}{12},\frac{-29}{72},
\frac{-1}{12},\frac{1}{144},1,\nonumber\\
&&\qquad\qquad\qquad\quad x_t)\Big]+\frac{1}{8}(C_{W^-\bar\chi^0_j \chi^+_i}^{L} C_{W^-\bar\chi^0_j \chi^+_i}^{R*}-C_{W^-\bar\chi^0_j \chi^+_i}^{R} C_{W^-\bar\chi^0_j \chi^+_i}^{L*})\frac{\partial^2 \rho_{2,1}}{\partial x_1^2}(x_W,x_t)\Big\},\nonumber\\
&&C_{8,NP}^{WW}(\mu_{EW})=\sum_{\chi^\pm_i,\chi^0_j}\frac{-3}{8\pi^2} \Big\{ (|C_{W^-\bar\chi^0_j \chi^+_i}^{L}|^2 +|C_{W^-\bar\chi^0_j \chi^+_i}^{R}|^2)P(\frac{-11}{12},\frac{-29}{72},\frac{-1}{12},\frac{1}{144},1,x_t)+\nonumber\\
&&\qquad\qquad\qquad\quad(C_{W^-\bar\chi^0_j \chi^+_i}^{R} C_{W^-\bar\chi^0_j \chi^+_i}^{L*}+C_{W^-\bar\chi^0_j \chi^+_i}^{L} C_{W^-\bar\chi^0_j \chi^+_i}^{R*})P(\frac{-1}{12},\frac{-5}{24},\frac{1}{12},\frac{-1}{144},1,x_t)+\nonumber\\
&&\qquad\qquad\qquad\quad(C_{W^-\bar\chi^0_j \chi^+_i}^{R} C_{W^-\bar\chi^0_j \chi^+_i}^{L*}-C_{W^-\bar\chi^0_j \chi^+_i}^{L} C_{W^-\bar\chi^0_j \chi^+_i}^{R*})P(\frac{1}{16},\frac{7}{24},0,0,1,x_t)\Big\},\nonumber\\
&&C_{7,NP}^{WH}(\mu_{EW})=\sum_{\chi^\pm_i,\chi^0_j}\frac{C_{H^-\bar d u}^{L}m_W^2}{4\sqrt{2}\pi^2m_dm_fV_{ud}^*}\Big\{\Big[\Re\Big(C_{H^-\bar\chi^0_j \chi^+_i}^{L} C_{W^-\bar\chi^0_j \chi^+_i}^{L}+C_{H^-\bar\chi^0_j \chi^+_i}^{R} C_{W^-\bar\chi^0_j \chi^+_i}^{R}\Big)-\nonumber\\
&&\qquad\qquad\qquad\quad i\Im\Big(C_{H^-\bar\chi^0_j \chi^+_i}^{L} C_{W^-\bar\chi^0_j \chi^+_i}^{L}+C_{H^-\bar\chi^0_j \chi^+_i}^{R} C_{W^-\bar\chi^0_j \chi^+_i}^{R}\Big)\Big]\Big(\frac{21}{64}-\frac{5}{288}+\frac{5}{24}J(m_W^2,\nonumber\\
&&\qquad\qquad\qquad\quad M_{H^\pm}^2,m_t^2)\Big)+\Big[\Re\Big(C_{H^-\bar\chi^0_j \chi^+_i}^{L} C_{W^-\bar\chi^0_j \chi^+_i}^{R}+C_{H^-\bar\chi^0_j \chi^+_i}^{R} C_{W^-\bar\chi^0_j \chi^+_i}^{L}\Big)-\nonumber\\
&&\qquad\qquad\qquad\quad i\Im\Big(C_{H^-\bar\chi^0_j \chi^+_i}^{L} C_{W^-\bar\chi^0_j \chi^+_i}^{R}+C_{H^-\bar\chi^0_j \chi^+_i}^{R} C_{W^-\bar\chi^0_j \chi^+_i}^{L}\Big)\Big]\Big(\frac{-1}{144}-\frac{1}{24}J(m_W^2,\nonumber\\
&&\qquad\qquad\qquad\quad M_{H^\pm}^2,m_t^2)\Big)-\Big[\Re\Big(C_{H^-\bar\chi^0_j \chi^+_i}^{L} C_{W^-\bar\chi^0_j \chi^+_i}^{L}-C_{H^-\bar\chi^0_j \chi^+_i}^{R} C_{W^-\bar\chi^0_j \chi^+_i}^{R}\Big)-\nonumber\\
&&\qquad\qquad\qquad\quad i\Im\Big(C_{H^-\bar\chi^0_j \chi^+_i}^{L} C_{W^-\bar\chi^0_j \chi^+_i}^{L}-C_{H^-\bar\chi^0_j \chi^+_i}^{R} C_{W^-\bar\chi^0_j \chi^+_i}^{R}\Big)\Big]\Big(\frac{16}{144}+\frac{1}{6}J(m_W^2,\nonumber\\
&&\qquad\qquad\qquad\quad M_{H^\pm}^2,m_t^2)\Big)-\Big[\Re\Big(C_{H^-\bar\chi^0_j \chi^+_i}^{L} C_{W^-\bar\chi^0_j \chi^+_i}^{R}-C_{H^-\bar\chi^0_j \chi^+_i}^{R} C_{W^-\bar\chi^0_j \chi^+_i}^{L}\Big)-\nonumber\\
&&\qquad\qquad\qquad\quad i\Im\Big(C_{H^-\bar\chi^0_j \chi^+_i}^{L} C_{W^-\bar\chi^0_j \chi^+_i}^{R}-C_{H^-\bar\chi^0_j \chi^+_i}^{R} C_{W^-\bar\chi^0_j \chi^+_i}^{L}\Big)\Big]\Big(\frac{1}{72}+\frac{1}{12}J(m_W^2,\nonumber\\
&&\qquad\qquad\qquad\quad M_{H^\pm}^2,m_t^2)\Big)\Big\},\nonumber\\
&&C_{8,NP}^{WH}(\mu_{EW})=\sum_{\chi^\pm_i,\chi^0_j}\frac{C_{H^-\bar d u}^{L}m_W^2}{4\sqrt{2}\pi^2m_dm_fV_{ud}^*}\Big\{\Big[\Re\Big(C_{H^-\bar\chi^0_j \chi^+_i}^{L} C_{W^-\bar\chi^0_j \chi^+_i}^{L}+C_{H^-\bar\chi^0_j \chi^+_i}^{R} C_{W^-\bar\chi^0_j \chi^+_i}^{R}\Big)-\nonumber\\
&&\qquad\qquad\qquad\quad i\Im\Big(C_{H^-\bar\chi^0_j \chi^+_i}^{L} C_{W^-\bar\chi^0_j \chi^+_i}^{L}+C_{H^-\bar\chi^0_j \chi^+_i}^{R} C_{W^-\bar\chi^0_j \chi^+_i}^{R}\Big)\Big]\Big(\frac{-1}{8\sqrt{2}}J(m_W^2,M_{H^\pm}^2,m_t^2)\Big)+\nonumber\\
&&\qquad\qquad\qquad\quad \Big[\Re\Big(C_{H^-\bar\chi^0_j \chi^+_i}^{L} C_{W^-\bar\chi^0_j \chi^+_i}^{R}+C_{H^-\bar\chi^0_j \chi^+_i}^{R} C_{W^-\bar\chi^0_j \chi^+_i}^{L}\Big)+\nonumber\\
&&\qquad\qquad\qquad\quad i\Im\Big(C_{H^-\bar\chi^0_j \chi^+_i}^{L} C_{W^-\bar\chi^0_j \chi^+_i}^{R}+C_{H^-\bar\chi^0_j \chi^+_i}^{R} C_{W^-\bar\chi^0_j \chi^+_i}^{L}\Big)\Big]\Big(\frac{-1}{8\sqrt{2}}J(m_W^2,M_{H^\pm}^2,m_t^2)\Big)+\nonumber\\
&&\qquad\qquad\qquad\quad\Big[\Re\Big(C_{H^-\bar\chi^0_j \chi^+_i}^{L} C_{W^-\bar\chi^0_j \chi^+_i}^{L}-C_{H^-\bar\chi^0_j \chi^+_i}^{R} C_{W^-\bar\chi^0_j \chi^+_i}^{R}\Big)-\nonumber\\
&&\qquad\qquad\qquad\quad i\Im\Big(C_{H^-\bar\chi^0_j \chi^+_i}^{L} C_{W^-\bar\chi^0_j \chi^+_i}^{L}-C_{H^-\bar\chi^0_j \chi^+_i}^{R} C_{W^-\bar\chi^0_j \chi^+_i}^{R}\Big)\Big]\Big(\frac{-1}{4\sqrt{2}}J(m_W^2,M_{H^\pm}^2,m_t^2)\Big)+\nonumber\\
&&\qquad\qquad\qquad\quad\Big[\Re\Big(C_{H^-\bar\chi^0_j \chi^+_i}^{L} C_{W^-\bar\chi^0_j \chi^+_i}^{R}+C_{H^-\bar\chi^0_j \chi^+_i}^{R} C_{W^-\bar\chi^0_j \chi^+_i}^{L}\Big)+\nonumber\\
&&\qquad\qquad\qquad\quad i\Im\Big(C_{H^-\bar\chi^0_j \chi^+_i}^{L} C_{W^-\bar\chi^0_j \chi^+_i}^{R}+C_{H^-\bar\chi^0_j \chi^+_i}^{R} C_{W^-\bar\chi^0_j \chi^+_i}^{L}\Big)\Big]\Big(\frac{1}{4\sqrt{2}}J(m_W^2,M_{H^\pm}^2,m_t^2)\Big)\Big\},\label{Bazzbsr}\nonumber\\
\end{eqnarray}
where $m_F$ runs all $m_{\chi_i^\pm}$, $m_{\chi_j^0}$, and
\begin{eqnarray}
&&\rho_{i,j}(x_1,x_2)=\frac{x_1^i \ln^j x_1-x_2^i \ln^j x_2}{x_1-x_2},\nonumber\\
&&P(y_1,y_2,y_3,y_4,x_1,x_2)=y_1\frac{\partial \rho_{1,1}(x_1,x_2)}{\partial x_1}+y_2\frac{\partial^2 \rho_{2,1}(x_1,x_2)}{\partial x_1^2}+\nonumber\\
&&\qquad\qquad\qquad\qquad\qquad\quad y_3\frac{\partial^3 \rho_{3,1}(x_1,x_2)}{\partial x_1^3}+y_4\frac{\partial^4 \rho_{4,1}(x_1,x_2)}{\partial x_1^4},\nonumber\\
&&J(x_1,x_2,x_3)=\ln m_F^2-\frac{\rho_{2,1}(x_1,x_3)-\rho_{2,2}(x_2,x_3)}{x_1^2-x_2^2}
\end{eqnarray}

\section{The Wilson coefficients of the process $B_s^0\rightarrow \mu^+\mu^-$. \label{wilsonbmumu}}
The Wilson coefficients corresponding to $b\rightarrow s\mu^+\mu^-$ can be written as
\begin{eqnarray}
&&C_{_{S,NP}}^{(1)}(\mu_{_{\rm EW}})=\sum_{_{\tilde U_i,\chi^-_j,\chi^-_k,S=h_l,A_l
}}\frac{C_{\mu^-S\mu^+}^L+C_{\mu^-S\mu^+}^R}{2(m_b^2-m_S^2)}\Big[C_{\tilde U_i\bar s \chi^-_j}^R C_{\bar \chi^-_j S \chi^-_k}^L C_{\bar \chi^-_k s \tilde U_i}^R G_2(x_{\tilde U_i},x_{\chi^\pm_j},x_{\chi^\pm_k})\nonumber\\
&&\qquad\qquad\qquad+M_{\chi^\pm_j}M_{\chi^\pm_k} C_{\tilde U_i\bar s \chi^-_j}^R C_{\bar \chi^-_j S \chi^-_k}^R C_{\bar \chi^-_k s \tilde U_i}^R G_1(x_{\tilde U_i},x_{\chi^\pm_j},x_{\chi^\pm_k})\Big]\nonumber\\
&&\qquad\qquad\qquad+\sum_{_{H^-_i,u_j,u_k,S=h_l,A_l}}\frac{C_{\mu^-S\mu^+}^L+C_{\mu^-S\mu^+}^R}{2(m_b^2-m_S^2)}\Big[C_{H^-_i\bar s u_j}^R C_{\bar u_j S u_k}^L C_{\bar u_k b H^-_i}^R G_2(x_{\tilde H^\pm_i},x_{u_j},x_{u_k})\nonumber\\
&&\qquad\qquad\qquad+m_{u_j}m_{u_k}C_{H^-_i\bar s u_j}^R C_{\bar u_j S u_k}^R C_{\bar u_k b H^-_i}^R G_1(x_{\tilde H^\pm_i},x_{u_j},x_{u_k})\Big],\nonumber\\
&&C_{_{P,NP}}^{(1)}(\mu_{_{\rm EW}})=\sum_{_{\tilde U_i,\chi^-_j,\chi^-_k,S=h_l,A_l
}}\frac{-C_{\mu^-S\mu^+}^L+C_{\mu^-S\mu^+}^R}{2(m_b^2-m_S^2)}\Big[C_{\tilde U_i\bar s \chi^-_j}^R C_{\bar \chi^-_j S \chi^-_k}^L C_{\bar \chi^-_k s \tilde U_i}^R G_2(x_{\tilde U_i},x_{\chi^\pm_j},x_{\chi^\pm_k})\nonumber\\
&&\qquad\qquad\qquad+M_{\chi^\pm_j}M_{\chi^\pm_k} C_{\tilde U_i\bar s \chi^-_j}^R C_{\bar \chi^-_j S \chi^-_k}^R C_{\bar \chi^-_k s \tilde U_i}^R G_1(x_{\tilde U_i},x_{\chi^\pm_j},x_{\chi^\pm_k})\Big]\nonumber\\
&&\qquad\qquad\qquad+\sum_{_{H^-_i,u_j,u_k,S=h_l,A_l}}\frac{-C_{\mu^-S\mu^+}^L+C_{\mu^-S\mu^+}^R}{2(m_b^2-m_S^2)}\Big[C_{H^-_i\bar s u_j}^R C_{\bar u_j S u_k}^L C_{\bar u_k b H^-_i}^R G_2(x_{\tilde H^\pm_i},x_{u_j},x_{u_k})\nonumber\\
&&\qquad\qquad\qquad+m_{u_j}m_{u_k}C_{H^-_i\bar s u_j}^R C_{\bar u_j S u_k}^R C_{\bar u_k b H^-_i}^R G_1(x_{\tilde H^\pm_i},x_{u_j},x_{u_k})\Big],
\end{eqnarray}
\begin{eqnarray}
&&C_{_{S,NP}}^{(2)}(\mu_{_{\rm EW}})=\sum_{u_i,H^\pm_j,H^\pm_k,S=h_l,A_l}\frac{1}{2(m_b^2-m_S^2)}m_{u_i}C_{\bar s u_i H^\pm_j}^RC_{\bar u_i b H^\pm_k}^RC_{S H^\pm_j H^\pm_k}G_1(x_{u_i},x_{H^\pm_j},x_{H^\pm_k})\nonumber\\
&&\qquad\qquad\qquad (C_{\mu^-S\mu^+}^L+C_{\mu^-S\mu^+}^R)\nonumber\\
&&\qquad\qquad\qquad+\sum_{\chi^\pm_i,\tilde U_j,\tilde U_k,S=h_l,A_l}\frac{1}{2(m_b^2-m_S^2)}m_{\chi^\pm_i}C_{\bar s \chi^\pm_i \tilde U_j}^RC_{\bar \chi^\pm_i b \tilde U_k}^RC_{S \tilde U_j \tilde U_k}G_1(x_{\chi^\pm_i},x_{\tilde U_j},x_{\tilde U_k})\nonumber\\
&&\qquad\qquad\qquad (C_{\mu^-S\mu^+}^L+C_{\mu^-S\mu^+}^R),\nonumber\\
&&C_{_{p,NP}}^{(2)}(\mu_{_{\rm EW}})=\sum_{u_i,H^\pm_j,H^\pm_k,S=h_l,A_l}\frac{1}{2(m_b^2-m_S^2)}m_{u_i}C_{\bar s u_i H^\pm_j}^RC_{\bar u_i b H^\pm_k}^RC_{S H^\pm_j H^\pm_k}G_1(x_{u_i},x_{H^\pm_j},x_{H^\pm_k})\nonumber\\
&&\qquad\qquad\qquad (-C_{\mu^-S\mu^+}^L+C_{\mu^-S\mu^+}^R)\nonumber\\
&&\qquad\qquad\qquad+\sum_{\chi^\pm_i,\tilde U_j,\tilde U_k,S=h_l,A_l}\frac{1}{2(m_b^2-m_S^2)}m_{\chi^\pm_i}C_{\bar s \chi^\pm_i \tilde U_j}^RC_{\bar \chi^\pm_i b \tilde U_k}^RC_{S \tilde U_j \tilde U_k}G_1(x_{\chi^\pm_i},x_{\tilde U_j},x_{\tilde U_k})\nonumber\\
&&\qquad\qquad\qquad (-C_{\mu^-S\mu^+}^L+C_{\mu^-S\mu^+}^R),
\end{eqnarray}
\begin{eqnarray}
&&C_{_{S,NP}}^{(3)}(\mu_{_{\rm EW}})=\sum_{u_i,H^\pm_k,S=h_l,A_l}\frac{-C_{W^\pm S H^\pm_k}}{2(m_b^2-m_S^2)}\Big[C_{\bar s W^\pm u_i}^LC_{\bar u_i H^\pm_k b}^RG_2(x_{u_i},1,x_{H^\pm_k})-2m_bm_{u_i}C_{\bar s W^\pm u_i}^L\nonumber\\
&&\qquad\qquad\qquad C_{\bar u_i H^\pm_k b}^LG_1(x_{u_i},1,x_{H^\pm_k})\Big](C_{\mu^-S\mu^+}^L+C_{\mu^-S\mu^+}^R),\nonumber\\
&&C_{_{P,NP}}^{(3)}(\mu_{_{\rm EW}})=\sum_{u_i,H^\pm_k,S=h_l,A_l}\frac{-C_{W^\pm S H^\pm_k}}{2(m_b^2-m_S^2)}\Big[C_{\bar s W^\pm u_i}^LC_{\bar u_i H^\pm_k b}^RG_2(x_{u_i},1,x_{H^\pm_k})-2m_bm_{u_i}C_{\bar s W^\pm u_i}^L\nonumber\\
&&\qquad\qquad\qquad C_{\bar u_i H^\pm_k b}^LG_1(x_{u_i},1,x_{H^\pm_k})\Big](-C_{\mu^-S\mu^+}^L+C_{\mu^-S\mu^+}^R),
\end{eqnarray}
\begin{eqnarray}
&&C_{_{S,NP}}^{(4)}(\mu_{_{\rm EW}})=\sum_{u_i,H^\pm_j,S=h_l,A_l}\frac{-C_{W^\pm S H^\pm_j}}{2(m_b^2-m_S^2)}C_{\bar s H^\pm_j u_i}^RC_{\bar u_i W^\pm b}^RG_2(x_{u_i},x_{H^\pm_j},1)(C_{\mu^-S\mu^+}^L+C_{\mu^-S\mu^+}^R),\nonumber\\
&&C_{_{S,NP}}^{(4)}(\mu_{_{\rm EW}})=\sum_{u_i,H^\pm_j,S=h_l,A_l}\frac{-C_{W^\pm S H^\pm_j}}{2(m_b^2-m_S^2)}C_{\bar s H^\pm_j u_i}^RC_{\bar u_i W^\pm b}^RG_2(x_{u_i},x_{H^\pm_j},1)(-C_{\mu^-S\mu^+}^L+C_{\mu^-S\mu^+}^R),\nonumber\\
\end{eqnarray}
\begin{eqnarray}
&&C_{_{9,NP}}^{(5)}(\mu_{_{\rm EW}})=\sum_{_{\tilde U_i,\chi^-_j,\chi^-_k,V
}}\frac{C_{\mu^-V\mu^+}^L+C_{\mu^-V\mu^+}^R}{-2(m_b^2-m_V^2)}\Big[-\frac{1}{2}C_{\tilde U_i\bar s \chi^-_j}^R C_{\bar \chi^-_j V \chi^-_k}^R C_{\bar \chi^-_k s \tilde U_i}^L G_2(x_{\tilde U_i},x_{\chi^\pm_j},x_{\chi^\pm_k})\nonumber\\
&&\qquad\qquad\qquad+M_{\chi^\pm_j}M_{\chi^\pm_k} C_{\tilde U_i\bar s \chi^-_j}^R C_{\bar \chi^-_j V \chi^-_k}^L C_{\bar \chi^-_k s \tilde U_i}^L G_1(x_{\tilde U_i},x_{\chi^\pm_j},x_{\chi^\pm_k})\Big]\nonumber\\
&&\qquad\qquad\qquad+\sum_{_{\tilde H^\pm_i,u_j,u_k,V
}}\frac{C_{\mu^-V\mu^+}^L+C_{\mu^-V\mu^+}^R}{-2(m_b^2-m_V^2)}\Big[-\frac{1}{2}C_{H^\pm_i\bar s u_j}^R C_{\bar u_j V u_k}^R C_{u_k s H^\pm_i}^L G_2(x_{H^\pm_i},x_{u_j},x_{u_k})\nonumber\\
&&\qquad\qquad\qquad+m_{u_j}m_{u_k} C_{H^\pm_i\bar s u_j}^R C_{\bar u_j V u_k}^L C_{\bar u_k s H^\pm_i}^L G_1(x_{H^\pm_i},x_{u_j},x_{u_k})\Big],\nonumber\\
&&C_{_{10,NP}}^{(5)}(\mu_{_{\rm EW}})=\sum_{_{\tilde U_i,\chi^-_j,\chi^-_k,V
}}\frac{-C_{\mu^-V\mu^+}^L+C_{\mu^-V\mu^+}^R}{-2(m_b^2-m_V^2)}\Big[-\frac{1}{2}C_{\tilde U_i\bar s \chi^-_j}^R C_{\bar \chi^-_j V \chi^-_k}^R C_{\bar \chi^-_k s \tilde U_i}^L G_2(x_{\tilde U_i},x_{\chi^\pm_j},x_{\chi^\pm_k})\nonumber\\
&&\qquad\qquad\qquad+M_{\chi^\pm_j}M_{\chi^\pm_k} C_{\tilde U_i\bar s \chi^-_j}^R C_{\bar \chi^-_j V \chi^-_k}^L C_{\bar \chi^-_k s \tilde U_i}^L G_1(x_{\tilde U_i},x_{\chi^\pm_j},x_{\chi^\pm_k})\Big]\nonumber\\
&&\qquad\qquad\qquad+\sum_{_{\tilde H^\pm_i,u_j,u_k,V
}}\frac{-C_{\mu^-V\mu^+}^L+C_{\mu^-V\mu^+}^R}{-2(m_b^2-m_V^2)}\Big[-\frac{1}{2}C_{H^\pm_i\bar s u_j}^R C_{\bar u_j V u_k}^R C_{u_k s H^\pm_i}^L G_2(x_{H^\pm_i},x_{u_j},x_{u_k})\nonumber\\
&&\qquad\qquad\qquad+m_{u_j}m_{u_k} C_{H^\pm_i\bar s u_j}^R C_{\bar u_j V u_k}^L C_{\bar u_k s H^\pm_i}^L G_1(x_{H^\pm_i},x_{u_j},x_{u_k})\Big],
\end{eqnarray}
\begin{eqnarray}
&&C_{_{9,NP}}^{(6)}(\mu_{_{\rm EW}})=\sum_{_{u_i,H^\pm_j,H^\pm_k,V
}}\frac{C_{\mu^-V\mu^+}^L+C_{\mu^-V\mu^+}^R}{4(m_b^2-m_V^2)}C_{\bar s u_i H^\pm_j}^RC_{\bar u_i b H^\pm_k}^LC_{V H^\pm_j H^\pm_k}G_2(x_{u_i},x_{H^\pm_j},x_{H^\pm_k})\nonumber\\
&&\qquad\qquad\qquad+\sum_{_{\chi^\pm_i,\tilde U_j,\tilde U_k,V
}}\frac{C_{\mu^-V\mu^+}^L+C_{\mu^-V\mu^+}^R}{4(m_b^2-m_V^2)}C_{\bar s \chi^\pm_i \tilde U_j}^RC_{\bar \chi^\pm_i b \tilde U_k}^LC_{V \tilde U_j \tilde U_k}G_2(x_{\chi^\pm_i},x_{\tilde U_j},x_{\tilde U_k}),\nonumber\\
&&C_{_{10,NP}}^{(6)}(\mu_{_{\rm EW}})=\sum_{_{u_i,H^\pm_j,H^\pm_k,V
}}\frac{-C_{\mu^-V\mu^+}^L+C_{\mu^-V\mu^+}^R}{4(m_b^2-m_V^2)}C_{\bar s u_i H^\pm_j}^RC_{\bar u_i b H^\pm_k}^LC_{V H^\pm_j H^\pm_k}G_2(x_{u_i},x_{H^\pm_j},x_{H^\pm_k})\nonumber\\
&&\qquad\qquad\qquad+\sum_{_{\chi^\pm_i,\tilde U_j,\tilde U_k,V
}}\frac{-C_{\mu^-V\mu^+}^L+C_{\mu^-V\mu^+}^R}{4(m_b^2-m_V^2)}C_{\bar s \chi^\pm_i \tilde U_j}^RC_{\bar \chi^\pm_i b \tilde U_k}^LC_{V \tilde U_j \tilde U_k}G_2(x_{\chi^\pm_i},x_{\tilde U_j},x_{\tilde U_k}),\nonumber\\
&&C_{_{S,NP}}^{(6)}(\mu_{_{\rm EW}})=\sum_{_{u_i,H^\pm_j,H^\pm_k,V
}}\frac{C_{\mu^-V\mu^+}^L+C_{\mu^-V\mu^+}^R}{-2(m_b^2-m_V^2)}m_bm_{u_i}C_{\bar s u_i H^\pm_j}^RC_{\bar u_i b H^\pm_k}^RC_{V H^\pm_j H^\pm_k}G_1(x_{u_i},x_{H^\pm_j},x_{H^\pm_k})\nonumber\\
&&\qquad\qquad\qquad+\sum_{_{\chi^\pm_i,\tilde U_j,\tilde U_k,V
}}\frac{C_{\mu^-V\mu^+}^L+C_{\mu^-V\mu^+}^R}{-2(m_b^2-m_V^2)}m_bm_{\chi^\pm_i}C_{\bar s \chi^\pm_i \tilde U_j}^RC_{\bar \chi^\pm_i b \tilde U_k}^RC_{V \tilde U_j \tilde U_k}G_1(x_{\chi^\pm_i},x_{\tilde U_j},x_{\tilde U_k}),\nonumber\\
&&C_{_{P,NP}}^{(6)}(\mu_{_{\rm EW}})=\sum_{_{u_i,H^\pm_j,H^\pm_k,V
}}\frac{C_{\mu^-V\mu^+}^L-C_{\mu^-V\mu^+}^R}{-2(m_b^2-m_V^2)}m_bm_{u_i}C_{\bar s u_i H^\pm_j}^RC_{\bar u_i b H^\pm_k}^RC_{V H^\pm_j H^\pm_k}G_1(x_{u_i},x_{H^\pm_j},x_{H^\pm_k})\nonumber\\
&&\qquad\qquad\qquad+\sum_{_{\chi^\pm_i,\tilde U_j,\tilde U_k,V
}}\frac{C_{\mu^-V\mu^+}^L-C_{\mu^-V\mu^+}^R}{-2(m_b^2-m_V^2)}m_bm_{\chi^\pm_i}C_{\bar s \chi^\pm_i \tilde U_j}^RC_{\bar \chi^\pm_i b \tilde U_k}^RC_{V \tilde U_j \tilde U_k}G_1(x_{\chi^\pm_i},x_{\tilde U_j},x_{\tilde U_k}),\nonumber\\
\end{eqnarray}
\begin{eqnarray}
&&C_{_{9,NP}}^{(7)}(\mu_{_{\rm EW}})=\sum_{_{u_i,H^\pm_k,V
}}\frac{C_{\mu^-V\mu^+}^L+C_{\mu^-V\mu^+}^R}{2(m_b^2-m_V^2)}m_{u_i}C_{\bar s u_i W^\pm}^LC_{\bar u_i b H^\pm_k}^LC_{V W^\pm H^\pm_k}G_1(x_{u_i},x_W,x_{H^\pm_k}),\nonumber\\
&&C_{_{10,NP}}^{(7)}(\mu_{_{\rm EW}})=\sum_{_{u_i,H^\pm_k,V
}}\frac{-C_{\mu^-V\mu^+}^L+C_{\mu^-V\mu^+}^R}{2(m_b^2-m_V^2)}m_{u_i}C_{\bar s u_i W^\pm}^LC_{\bar u_i b H^\pm_k}^LC_{V W^\pm H^\pm_k}G_1(x_{u_i},x_W,x_{H^\pm_k}),\nonumber\\
\end{eqnarray}
\begin{eqnarray}
&&C_{_{9,NP}}^{(8)}(\mu_{_{\rm EW}})=\sum_{_{u_i,H^\pm_j,V
}}\frac{C_{\mu^-V\mu^+}^L+C_{\mu^-V\mu^+}^R}{2(m_b^2-m_V^2)}m_{u_i}C_{\bar s u_i H^\pm_j}^RC_{\bar u_i b W^\pm}^LC_{V W^\pm H^\pm_k}G_1(x_{u_i},x_{H^\pm_j},x_W),\nonumber\\
&&C_{_{10,NP}}^{(8)}(\mu_{_{\rm EW}})=\sum_{_{u_i,H^\pm_j,V
}}\frac{-C_{\mu^-V\mu^+}^L+C_{\mu^-V\mu^+}^R}{2(m_b^2-m_V^2)}m_{u_i}C_{\bar s u_i H^\pm_j}^RC_{\bar u_i b W^\pm}^LC_{V W^\pm H^\pm_k}G_1(x_{u_i},x_{H^\pm_j},x_W),\nonumber\\
\end{eqnarray}
\begin{eqnarray}
&&C_{_{9,NP}}^{(9)}(\mu_{_{\rm EW}})=\sum_{_{\tilde U_i,\chi^\pm_j,\chi^\pm_k,\tilde\nu_l
}}-\frac{1}{8}C_{\bar s \tilde U_i \chi^\pm_j}^RC_{\bar \chi^\pm_j \mu^+ \tilde\nu_l}^L(C_{\bar\mu^-\tilde\nu_l\chi^\pm_k}^LC_{\bar\chi^\pm_k \tilde U_i b}^R+C_{\bar\mu^-\tilde\nu_l\chi^\pm_k}^RC_{\bar\chi^\pm_k \tilde U_i b}^L)\nonumber\\
&&\qquad\qquad\qquad G_4(x_{\tilde U_i},x_{\chi^\pm_j},x_{\chi^\pm_k},x_{\tilde \nu_l}),\nonumber\\
&&C_{_{10,NP}}^{(9)}(\mu_{_{\rm EW}})=\sum_{_{\tilde U_i,\chi^\pm_j,\chi^\pm_k,\tilde\nu_l
}}-\frac{1}{8}C_{\bar s \tilde U_i \chi^\pm_j}^RC_{\bar \chi^\pm_j \mu^+ \tilde\nu_l}^L(C_{\bar\mu^-\tilde\nu_l\chi^\pm_k}^LC_{\bar\chi^\pm_k \tilde U_i b}^R-C_{\bar\mu^-\tilde\nu_l\chi^\pm_k}^RC_{\bar\chi^\pm_k \tilde U_i b}^L)\nonumber\\
&&\qquad\qquad\qquad G_4(x_{\tilde U_i},x_{\chi^\pm_j},x_{\chi^\pm_k},x_{\tilde \nu_l}),\nonumber\\
&&C_{_{S,NP}}^{(9)}(\mu_{_{\rm EW}})=\sum_{_{\tilde U_i,\chi^\pm_j,\chi^\pm_k,\tilde\nu_l
}}-\frac{1}{2}M_{\chi^\pm_j}M_{\chi^\pm_k}C_{\bar s \tilde U_i \chi^\pm_j}^RC_{\bar \chi^\pm_j \mu^+ \tilde\nu_l}^R(C_{\bar\mu^-\tilde\nu_l\chi^\pm_k}^LC_{\bar\chi^\pm_k \tilde U_i b}^L+C_{\bar\mu^-\tilde\nu_l\chi^\pm_k}^RC_{\bar\chi^\pm_k \tilde U_i b}^R)\nonumber\\
&&\qquad\qquad\qquad G_3(x_{\tilde U_i},x_{\chi^\pm_j},x_{\chi^\pm_k},x_{\tilde \nu_l}),\nonumber\\
&&C_{_{P,NP}}^{(9)}(\mu_{_{\rm EW}})=\sum_{_{\tilde U_i,\chi^\pm_j,\chi^\pm_k,\tilde\nu_l
}}-\frac{1}{2}M_{\chi^\pm_j}M_{\chi^\pm_k}C_{\bar s \tilde U_i \chi^\pm_j}^RC_{\bar \chi^\pm_j \mu^+ \tilde\nu_l}^R(-C_{\bar\mu^-\tilde\nu_l\chi^\pm_k}^LC_{\bar\chi^\pm_k \tilde U_i b}^L+C_{\bar\mu^-\tilde\nu_l\chi^\pm_k}^RC_{\bar\chi^\pm_k \tilde U_i b}^R)\nonumber\\
&&\qquad\qquad\qquad G_3(x_{\tilde U_i},x_{\chi^\pm_j},x_{\chi^\pm_k},x_{\tilde \nu_l}),
\end{eqnarray}
\begin{eqnarray}
&&C_{_{9,NP}}^{(10)}(\mu_{_{\rm EW}})=\sum_{_{u_i,H^\pm_j,H^\pm_k,\tilde\nu_l}}\frac{1}{8}C_{\bar s u_i H^\pm_j}^RC_{\bar u_i b H^\pm_k}^L(C_{\bar\mu^-H^\pm_k\nu_l}^LC_{\bar\nu_l H^\pm_j \mu^+}^R+C_{\bar\mu^-H^\pm_k\nu_l}^RC_{\bar\nu_l H^\pm_j \mu^+}^L)\nonumber\\
&&\qquad\qquad\qquad G_4(x_{u_i},x_{H^\pm_j},x_{H^\pm_k},x_{\nu_l}),\nonumber\\
&&C_{_{10,NP}}^{(10)}(\mu_{_{\rm EW}})=\sum_{_{u_i,H^\pm_j,H^\pm_k,\tilde\nu_l}}\frac{1}{8}C_{\bar s u_i H^\pm_j}^RC_{\bar u_i b H^\pm_k}^L(C_{\bar\mu^-H^\pm_k\nu_l}^LC_{\bar\nu_l H^\pm_j \mu^+}^R-C_{\bar\mu^-H^\pm_k\nu_l}^RC_{\bar\nu_l H^\pm_j \mu^+}^L)\nonumber\\
&&\qquad\qquad\qquad G_4(x_{u_i},x_{H^\pm_j},x_{H^\pm_k},x_{\nu_l}),
\end{eqnarray}
\begin{eqnarray}
&&C_{_{S,NP}}^{(11)}(\mu_{_{\rm EW}})=-\sum_{_{u_i,H^\pm_j,\nu_l}}\frac{1}{2}C_{\bar s u_i H^\pm_j}^RC_{\bar u_i b W^\pm}^RC_{\bar\mu^-W^\pm\nu_l}^RC_{\bar\nu_l H^\pm_j \mu^+}^L G_4(x_{u_i},x_{H^\pm_j},x_W,x_{\nu_l}),\nonumber\\
&&C_{_{P,NP}}^{(11)}(\mu_{_{\rm EW}})=-\sum_{_{u_i,H^\pm_j,\nu_l}}\frac{1}{2}C_{\bar s u_i H^\pm_j}^RC_{\bar u_i b W^\pm}^RC_{\bar\mu^-W^\pm\nu_l}^RC_{\bar\nu_l H^\pm_j \mu^+}^L G_4(x_{u_i},x_{H^\pm_j},x_W,x_{\nu_l}).
\end{eqnarray}
\begin{eqnarray}
&&C_{_{9,NP}}^{WW}(\mu_{_{\rm EW}})=\sum_{_{\chi^+_i,\chi^0_j,V}}\frac{(C_{\mu^-V\mu^+}^R+C_{\mu^-V\mu^+}^L)g_{s}^2}{128\pi^4 s_W^2}\frac{m_b^2}{m_b^2-m_V^2} \Big\{ \Big[(|C_{W^-\bar\chi^0_j \chi^+_i}^{L}|^2 +|C_{W^-\bar\chi^0_j \chi^+_i}^{R}|^2) \nonumber\\
&&\qquad\qquad\qquad\quad P(\frac{1}{8},\frac{1}{4},\frac{-1}{48},\frac{-1}{144},1,x_t)+(|C_{W^-\bar\chi^0_j \chi^+_i}^{L}|^2 -|C_{W^-\bar\chi^0_j \chi^+_i}^{R}|^2)\frac{\partial^1 \rho_{1,1}}{\partial x_1}(x_W,x_t)+\nonumber\\
&&\qquad\qquad\qquad\quad(C_{W^-\bar\chi^0_j \chi^+_i}^{R}C_{W^-\bar\chi^0_j \chi^+_i}^{L*}+C_{W^-\bar\chi^0_j \chi^+_i}^{L}C_{W^-\bar\chi^0_j \chi^+_i}^{R*}) P(\frac{1}{16},\frac{1}{4},\frac{-1}{16},\frac{1}{144},1,x_t)+\nonumber\\
&&\qquad\qquad\qquad\quad \frac{1}{8}(C_{W^-\bar\chi^0_j \chi^+_i}^{L} C_{W^-\bar\chi^0_j \chi^+_i}^{R*}-C_{W^-\bar\chi^0_j \chi^+_i}^{R} C_{W^-\bar\chi^0_j \chi^+_i}^{L*})\frac{\partial^2 \rho_{2,1}}{\partial x_1^2}(x_W,x_t)\Big]C_{VW^-W^-}+\nonumber\\
&&\qquad\qquad\qquad\quad\Big[(|C_{W^-\bar\chi^0_j \chi^+_i}^{L}|^2 +|C_{W^-\bar\chi^0_j \chi^+_i}^{R}|^2)P(\frac{1}{144},\frac{-1}{12},\frac{-29}{72},\frac{-11}{12},1,x_t)+(C_{W^-\bar\chi^0_j \chi^+_i}^{R}\nonumber\\
&&\qquad\qquad\qquad\quad C_{W^-\bar\chi^0_j \chi^+_i}^{L*}+C_{W^-\bar\chi^0_j \chi^+_i}^{L}C_{W^-\bar\chi^0_j \chi^+_i}^{R*}) P(\frac{-1}{144},\frac{1}{12},\frac{-5}{24},\frac{-1}{12},1,x_t)\Big]C_{\bar tVt}+\nonumber\\
&&\qquad\qquad\qquad\quad \Big[(|C_{W^-\bar\chi^0_j \chi^+_i}^{L}|^2 +|C_{W^-\bar\chi^0_j \chi^+_i}^{R}|^2)P(\frac{3}{16},\frac{3}{16},0,0,1,x_t)+\frac{1}{16}(C_{W^-\bar\chi^0_j \chi^+_i}^{L}\nonumber\\
&&\qquad\qquad\qquad\quad C_{W^-\bar\chi^0_j \chi^+_i}^{R*}-C_{W^-\bar\chi^0_j \chi^+_i}^{R} C_{W^-\bar\chi^0_j \chi^+_i}^{L*})\frac{\partial^2 \rho_{2,1}}{\partial x_1^2}(x_W,x_t)\Big](C_{\bar\chi_j^0V\chi_j^0}+C_{\bar\chi_i^+V\chi_i^+})\Big\},\nonumber\\
\end{eqnarray}
\begin{eqnarray}
&&C_{_{10,NP}}^{WW}(\mu_{_{\rm EW}})=\sum_{_{\chi^+_i,\chi^0_j,V}}\frac{(C_{\mu^-V\mu^+}^R-C_{\mu^-V\mu^+}^L)g_{s}^2}{128\pi^4 s_W^2}\frac{m_b^2}{m_b^2-m_V^2} \Big\{ \Big[(|C_{W^-\bar\chi^0_j \chi^+_i}^{L}|^2 +|C_{W^-\bar\chi^0_j \chi^+_i}^{R}|^2) \nonumber\\
&&\qquad\qquad\qquad\quad P(\frac{1}{8},\frac{1}{4},\frac{-1}{48},\frac{-1}{144},1,x_t)+(|C_{W^-\bar\chi^0_j \chi^+_i}^{L}|^2 -|C_{W^-\bar\chi^0_j \chi^+_i}^{R}|^2)\frac{\partial^1 \rho_{1,1}}{\partial x_1}(x_W,x_t)+\nonumber\\
&&\qquad\qquad\qquad\quad(C_{W^-\bar\chi^0_j \chi^+_i}^{R}C_{W^-\bar\chi^0_j \chi^+_i}^{L*}+C_{W^-\bar\chi^0_j \chi^+_i}^{L}C_{W^-\bar\chi^0_j \chi^+_i}^{R*}) P(\frac{1}{16},\frac{1}{4},\frac{-1}{16},\frac{1}{144},1,x_t)+\nonumber\\
&&\qquad\qquad\qquad\quad \frac{1}{8}(C_{W^-\bar\chi^0_j \chi^+_i}^{L} C_{W^-\bar\chi^0_j \chi^+_i}^{R*}-C_{W^-\bar\chi^0_j \chi^+_i}^{R} C_{W^-\bar\chi^0_j \chi^+_i}^{L*})\frac{\partial^2 \rho_{2,1}}{\partial x_1^2}(x_W,x_t)\Big]C_{VW^-W^-}+\nonumber\\
&&\qquad\qquad\qquad\quad\Big[(|C_{W^-\bar\chi^0_j \chi^+_i}^{L}|^2 +|C_{W^-\bar\chi^0_j \chi^+_i}^{R}|^2)P(\frac{1}{144},\frac{-1}{12},\frac{-29}{72},\frac{-11}{12},1,x_t)+(C_{W^-\bar\chi^0_j \chi^+_i}^{R}\nonumber\\
&&\qquad\qquad\qquad\quad C_{W^-\bar\chi^0_j \chi^+_i}^{L*}+C_{W^-\bar\chi^0_j \chi^+_i}^{L}C_{W^-\bar\chi^0_j \chi^+_i}^{R*}) P(\frac{-1}{144},\frac{1}{12},\frac{-5}{24},\frac{-1}{12},1,x_t)\Big]C_{\bar tVt}+\nonumber\\
&&\qquad\qquad\qquad\quad \Big[(|C_{W^-\bar\chi^0_j \chi^+_i}^{L}|^2 +|C_{W^-\bar\chi^0_j \chi^+_i}^{R}|^2)P(\frac{3}{16},\frac{3}{16},0,0,1,x_t)+\frac{1}{16}(C_{W^-\bar\chi^0_j \chi^+_i}^{L}\nonumber\\
&&\qquad\qquad\qquad\quad C_{W^-\bar\chi^0_j \chi^+_i}^{R*}-C_{W^-\bar\chi^0_j \chi^+_i}^{R} C_{W^-\bar\chi^0_j \chi^+_i}^{L*})\frac{\partial^2 \rho_{2,1}}{\partial x_1^2}(x_W,x_t)\Big](C_{\bar\chi_j^0V\chi_j^0}+C_{\bar\chi_i^+V\chi_i^+})\Big\},\nonumber\\
\end{eqnarray}
where $V$ denotes photon $\gamma$, $Z$ boson, $Z'$ boson. And $C_{\bar tVt},\;C_{\bar\chi_j^0V\chi_j^0},\;C_{\bar\chi_i^+V\chi_i^+}$ denote the vector parts of the corresponding interaction vertex. The concrete expressions for $G_k(k=1,...,4)$ can be given as:
\begin{eqnarray}
&&G_1(x_1,x_2,x_3)=\frac{-1}{m_W^2}\Big[\frac{x_1{\rm ln} x_1}{(x_2-x_1)(x_3-x_1)}+\frac{x_2{\rm ln} x_2}{(x_1-x_2)(x_3-x_2)}+\frac{x_3{\rm ln} x_3}{(x_1-x_3)(x_2-x_3)}\Big],\nonumber\\
&&G_2(x_1,x_2,x_3)=-\frac{x_1^2{\rm ln} x_1}{(x_2-x_1)(x_3-x_1)}-\frac{x_2^2{\rm ln} x_2}{(x_1-x_2)(x_3-x_2)}-\frac{x_3^2{\rm ln} x_3}{(x_1-x_3)(x_2-x_3)},\nonumber\\
&&G_3(x_1,x_2,x_3,x_4)=\frac{1}{m_W^4}\Big[\frac{x_1{\rm ln} x_1}{(x_2-x_1)(x_3-x_1)(x_4-x_1)}+\frac{x_2{\rm ln} x_2}{(x_1-x_2)(x_3-x_2)(x_4-x_2)}+\nonumber\\
&&\qquad\qquad\qquad\qquad\frac{x_3{\rm ln} x_3}{(x_1-x_3)(x_2-x_3)(x_4-x_3)}\frac{x_4{\rm ln} x_4}{(x_1-x_4)(x_2-x_4)(x_3-x_4)}\Big],\nonumber\\
&&G_4(x_1,x_2,x_3,x_4)=\frac{1}{m_W^2}\Big[\frac{x_1^2{\rm ln} x_1}{(x_2-x_1)(x_3-x_1)(x_4-x_1)}+\frac{x_2^2{\rm ln} x_2}{(x_1-x_2)(x_3-x_2)(x_4-x_2)}+\nonumber\\
&&\qquad\qquad\qquad\qquad\frac{x_3^2{\rm ln} x_3}{(x_1-x_3)(x_2-x_3)(x_4-x_3)}\frac{x_4^2{\rm ln} x_4}{(x_1-x_4)(x_2-x_4)(x_3-x_4)}\Big].\nonumber\\
\end{eqnarray}

\end{document}